\documentclass[12pt,aps,tightenlines]{revtex4}

\usepackage{amsmath}
\usepackage{amsfonts}
\usepackage{graphicx}
\usepackage{subfigure}

\graphicspath{{figs/}}

\newcommand{\Mag}{\bm{m}}
\newcommand{\Subm}{m}

\newcommand{\Int}{\int_{-\infty}^{\infty}}

\begin{document}
\title{Emergent singular solutions of non-local density-magnetization equations
in one dimension}
\author{Darryl D. Holm$^{1,2}$, Lennon \'O N\'araigh$^{1,}\!\!\!$
\footnote{Corresponding author. Email: lennon.o-naraigh@imperial.ac.uk}
, and Cesare Tronci$^{1,3}$}

\affiliation{\scriptsize $^1$Department of Mathematics, Imperial College
  London, SW7 2AZ, United Kingdom\\
$^2$Computer and Computational Science Division, Los Alamos National Laboratory, Los Alamos, NM, 87545 USA
\\
$^3$TERA Foundation for Oncological Hadrontherapy, 11 V. Puccini, Novara 28100, Italy
}
%
%
%
%
%
%
%
%
\date{\today}

\begin{abstract}
We investigate the emergence of singular solutions in a non-local model for
a magnetic system.  We study a modified
Gilbert-type equation for the magnetization vector and find that the evolution
depends strongly on the length scales of the non-local effects.  We pass
to
a coupled density-magnetization model and perform a linear stability analysis,
noting the effect of the length scales of non-locality on the system's stability
properties.  We carry out numerical simulations of the coupled system and
find that singular solutions emerge from smooth initial data.  The singular
solutions represent a collection of interacting particles (clumpons).  By
restricting ourselves to the two-clumpon case, we are reduced to a two-dimensional
dynamical system that is readily analyzed, and thus we classify the different
clumpon interactions possible.
\end{abstract}
%
%
%
\maketitle
\section{Introduction}
\label{sec:intro}
In recent years, the modeling of nanoscale physics has become important,
both because of industrial applications~\cite{Denis2002,Veinot2002,Moller2001,Wijnhoven1998},
and because of the development of experiments that probe these small scales~\cite{WolfBook}.
One particular problem is the modeling of aggregation, in which microscopic
particles collapse under the potential they exert on each other, and form
mesoscopic structures that in turn behave like particles.

In a series of papers, Holm, Putkaradze and Tronci~\cite{Darryl_eqn1,Darryl_eqn2_0,Darryl_eqn2_1,Darryl_eqn3,Darryl_eqn5,Darryl_eqn6}
have focused on the derivation of aggregation equations that
possess emergent singular solutions.  Continuum aggregation equations have
been used
to model gravitational collapse and the subsequent emergence of stars~\cite{ChandraStars},
the localization of biological populations~\cite{KellerSegel1970,Segel1985,Topaz2006},
and the self-assembly of nanoparticles~\cite{Putkaradze2005}.  These are
complexes of atoms or molecules that form mesoscale structures with particle-like
behavior.
The utility of the Holm--Putkaradze model lies in its emphasis on non-local
physics, and the emergence of singular solutions from smooth initial data.
 Because of the singular (delta-function) behavior of the model, it is an
 appropriate way to describe the universal phenomena of 
aggregation and the subsequent formation of particle-like structures.  Indeed
in this framework, it is possible to prescribe the dynamics of the particle-like
structures after collapse.
Thus, the model
provides a description of directed self-assembly in nanophysics~\cite{Xia2004,Putkaradze2005},
in which the detailed physics is less important than the effective medium
properties of the dynamics.

In this work we focus on equations introduced by Holm, Putkaradze and Tronci
for the aggregation of oriented particles~\cite{Darryl_eqn1,Darryl_eqn3}.
 We treat the
initial state of the system as a continuum, a good approximation in nanophysics
applications~\cite{Forest2007}.
 One realization of this problem is in nanomagnetics, in which particles
 with a definite magnetic moment collapse and form mesoscale structures,
 that in turn have a definite magnetic moment.  Thus, in this paper we refer
 to the orientation vector in our continuum picture as the \emph{magnetization}.
  We investigate these equations numerically and study their evolution and
  aggregation properties.  One aspect of non-local problems, already mentioned
  in~\cite{Darryl_eqn6}, is the effect of competition between the length
  scales of non-locality
  on the system evolution.  We shall highlight this effect with a linear
  stability analysis of the full density-magnetization equations.

This paper is organized as follows.  In Sec.~\ref{sec:gilbert} we introduce
a non-local Gilbert (NG) equation to describe non-local interactions in a
magnetic
system.  We investigate the competition between the system's two length scales
of non-locality.   In Sec.~\ref{sec:mag_dens} we
introduce a coupled density-magnetization system that generates singular
solutions.  We examine the competition of length scales through a linear
stability analysis and through the study of the dynamical equations for a
simple singular solution that describes the interaction of two particle-like
objects (clumpons).  We perform numerical simulations that
highlight the emergence of singular solutions from smooth initial data. 
We draw our conclusions in Sec.~\ref{sec:conclusions}.

\section{The Non-local Gilbert Equation}
\label{sec:gilbert}

In this section we study a magnetization equation that in form is similar
to the Gilbert equation, that is, the Landau--Lifshitz--Gilbert equation
in
the over-damped limit~\cite{GilbertIEEE,Weinan2000}.  The equation we focus
on incorporates
non-local effects, and was introduced in~\cite{Darryl_eqn1}.  We study the
evolution and energetics of this equation, and examine the importance of
the problem length scales in determining the evolution.

We study the following non-local Gilbert (NG) equation,
\begin{equation}
\frac{\partial\Mag}{\partial t} = \Mag\times\left(\bm{\mu}_\Subm\times\frac{\delta{E}}{\delta\Mag}\right),
\label{eq:mag_eqn}
\end{equation}
where $\Mag$ is the magnetization density, $\bm{\mu}_\Subm$ is the mobility,
defined as
\[
\bm{\mu}_\Subm = \left(1-\beta^2\partial_x^2\right)^{-1}\Mag,
\]
and $\delta E/\delta\Mag$ is the variational derivative of the energy,
\[
\frac{\delta E}{\delta\Mag} = \left(1-\alpha^2\partial_x^2\right)^{-1}\Mag.
\]
The smoothened magnetization $\bm{\mu}_\Subm$ and the force $\delta{E}/\delta{\bm{m}}$
can be computed using the theory of Green's functions.  In particular,
\[
\bm{\mu}_\Subm\left(x,t\right)=\int_\Omega{dy} H_{\beta}\left(x-y\right)\Mag\left(y,t\right):=H_{\beta}*\Mag\left(x,t\right).
\]
Here $*$ denotes the convolution of functions, and the kernel $H_{\beta}\left(x\right)$
satisfies the equation
\begin{equation}
\left(1-\beta^2\frac{d^2}{dx^2}\right)H_\beta\left(x\right)=\delta\left(x\right).
\label{eq:kernel}
\end{equation}
The function $\delta\left(x\right)$ is the Dirac delta function.  Equation~\eqref{eq:kernel}
is solved subject to conditions imposed on the boundary of the domain $\Omega$.
 In this paper we shall work with a periodic domain $\Omega=\left[-L/2,L/2\right]$
 or $\Omega=\left[0,L\right]$,
 although other boundary conditions are possible.  Note that Eq.~\eqref{eq:mag_eqn}
 has a family of non-trivial equilibrium states given by
\[
\Mag_{\mathrm{eq}}\left(x\right)=\Mag_0\sin\left(kx+\phi_0\right),
\]
where $\Mag_0$ is a constant vector, $k$ is some wave number, and $\phi_0$
is a constant phase.  The derivation of this solution is subject to the boundary
conditions discussed in Sec.~\ref{sec:mag_dens}.

By setting $\beta=0$ and replacing $\left(1-\alpha^2\partial_x^2\right)^{-1}$
with $-\partial_x^2$, we recover the more familiar Landau--Lifshitz--Gilbert
equation, in the overdamped limit~\cite{GilbertIEEE},
\begin{equation}
\frac{\partial\Mag}{\partial t} = -\Mag\times\left(\Mag\times\frac{\partial^2\Mag}{\partial{x}^2}\right).
\label{eq:gilbert}
\end{equation}
%
%
%
%
%
%
%
%
Equation~\eqref{eq:mag_eqn} possesses several features that will be useful
in understanding the numerical simulations.  There is an energy
functional
\begin{equation}
E\left(t\right)=\tfrac{1}{2}\int_\Omega{dx}\Mag\cdot\left(1-\alpha^2\partial_x^2\right)^{-1}\Mag,
\label{eq:energy}
\end{equation}
which evolves in time according to the relation
\begin{eqnarray}
\frac{dE}{dt}&=&\int_\Omega{dx}\left[\bm{\mu}_m\cdot\left(1-\alpha^2\partial_x^2\right)^{-1}\Mag\right]\left[\Mag\cdot\left(1-\alpha^2\partial_x^2\right)^{-1}\Mag\right]\nonumber\\
&\phantom{a}&\phantom{aaaaaaaaaaaaaaaaaaaaaadaaaa}
-\int_\Omega{dx}\left(\bm{\mu}_m\cdot\Mag\right)\left[\left(1-\alpha^2\partial_x^2\right)^{-1}\Mag\right]^2,\nonumber\\
&=&-\int_\Omega{dx}\left[\Mag\times\left(1-\alpha^2\partial_x^2\right)^{-1}\Mag\right]\cdot\left[\bm{\mu}_m\times\left(1-\alpha^2\partial_x^2\right)^{-1}\Mag\right].
\label{eq:dt_energy}
\end{eqnarray}
This is not necessarily a non-increasing function of time, although setting $\beta=0$
gives
\begin{eqnarray}
\left(\frac{dE}{dt}\right)_{\beta=0}&=&\int_\Omega{dx}\left[\Mag\cdot\left(1-\alpha^2\partial_x^2\right)^{-1}\Mag\right]^2-
\int_\Omega{dx}\Mag^2\left[\left(1-\alpha^2\partial_x^2\right)^{-1}\Mag\right]^2,\nonumber\\
&=&\int_\Omega{dx}\Mag^2\left[\left(1-\alpha^2\partial_x^2\right)^{-1}\Mag\right]^2\left(\cos^2\varphi-1\right)\leq0,
\label{eq:dt_energy_beta0}
\end{eqnarray}
where $\varphi$ is the angle between $\Mag$ and $\left(1-\alpha^2\partial_x^2\right)^{-1}\Mag$.
 In the special case when $\beta\rightarrow0$, we therefore expect $E\left(t\right)$
  to be a non-increasing function of time.  On the other hand, inspection
  of Eq.~\eqref{eq:dt_energy} shows that as $\alpha\rightarrow0$, the energy
  tends to a constant.
Additionally, the magnitude of the vector $\Mag$ is conserved.  This can
 be shown by multiplying Eq.~\eqref{eq:mag_eqn} by $\Mag$, and by exploiting
 the antisymmetry of the cross product.  Thus, we are interested only in
 the orientation of the vector $\Mag$; this can be parametrized by two
 angles on the sphere:
\begin{figure}
  \scalebox{0.4}[0.4]{\includegraphics*[viewport=0 0 420 330]{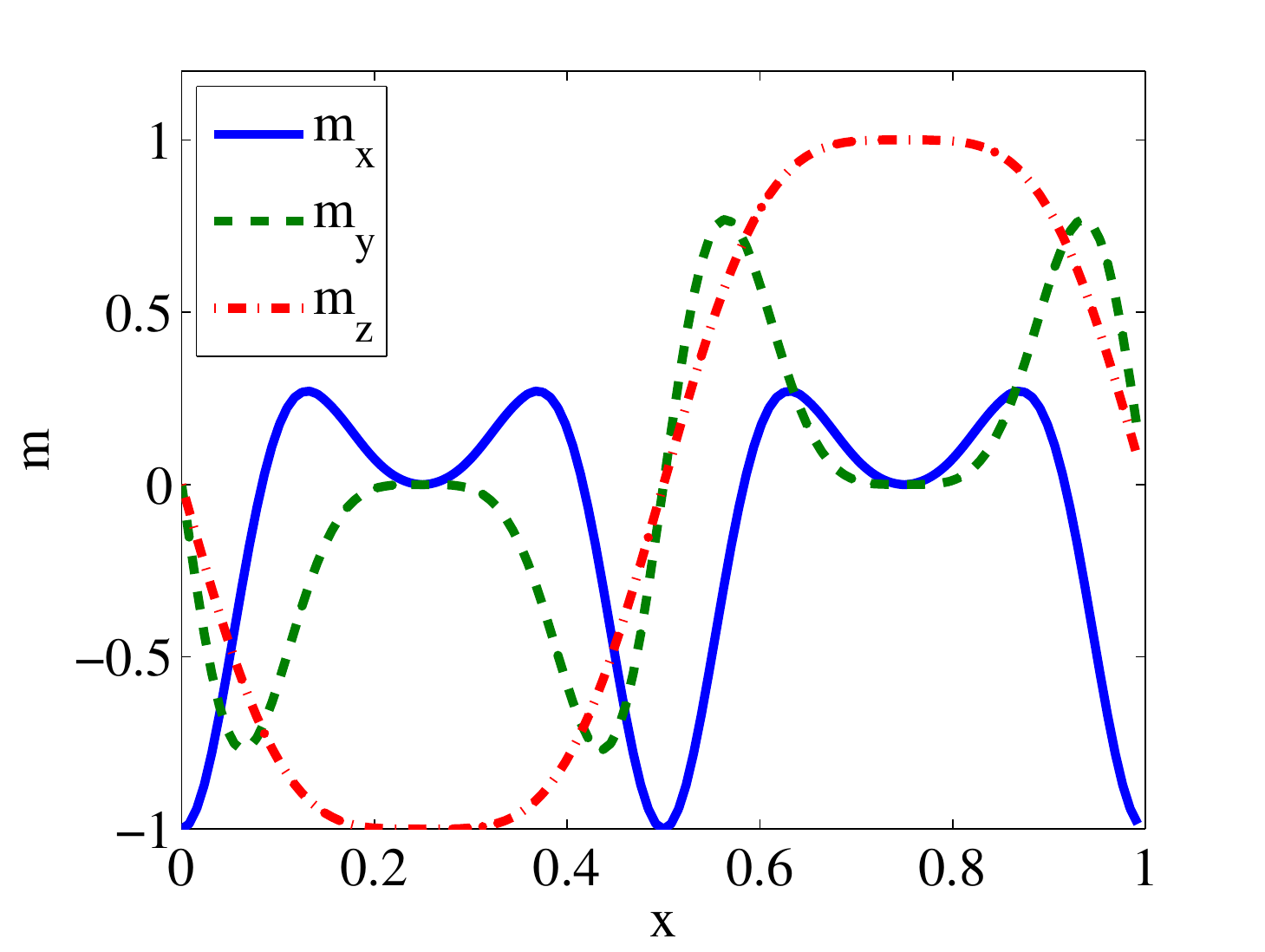}}
\caption{(Color online) The initial data for the magnetization equation~\eqref{eq:mag_eqn}.
 This initialization is obtained by allowing the orientation angles
 of the magnetization vector to vary sinusoidally in space, as in Eq.~\eqref{eq:initial}.
  Here the wave number of the variation is equal to the fundamental wave
  number
  $2\pi/L$.}
\label{fig:initial}
\end{figure}
the azimuthal angle $\theta\left(\bm{x},t\right)$, and the polar angle $\phi\left(\bm{x},t\right)$,
where
\begin{equation}
m_x=|\Mag|\cos\phi\sin\theta,\qquad
m_y=|\Mag|\sin\phi\sin\theta,\qquad
m_z=|\Mag|\cos\theta,
\label{eq:spherical_polars}
\end{equation}
and where $\phi\in\left[0,2\pi\right)$, and $\theta\in\left[0,\pi\right]$.

We carry out numerical simulations of Eqs.~\eqref{eq:mag_eqn} and~\eqref{eq:gilbert}
on a periodic domain $\left[0,L\right]$, and outline the findings in
what
follows.  Motivated by the change of coordinates~\eqref{eq:spherical_polars},
we choose the initial data
\begin{equation}
\phi_0\left(x\right)=\pi\left(1+\sin\left(2r\pi x/L\right)\right),\qquad
\theta_0\left(x\right)=\tfrac{1}{2}\pi\left(1+\sin\left(2\pi s x/L\right)\right),
\label{eq:initial}
\end{equation}
where $r$ and $s$ are integers.  These data are shown in Fig.~\ref{fig:initial}.

\begin{figure}[htb]
\subfigure[]{
  \scalebox{0.3}[0.3]{\includegraphics*[viewport=0 0 420 330]{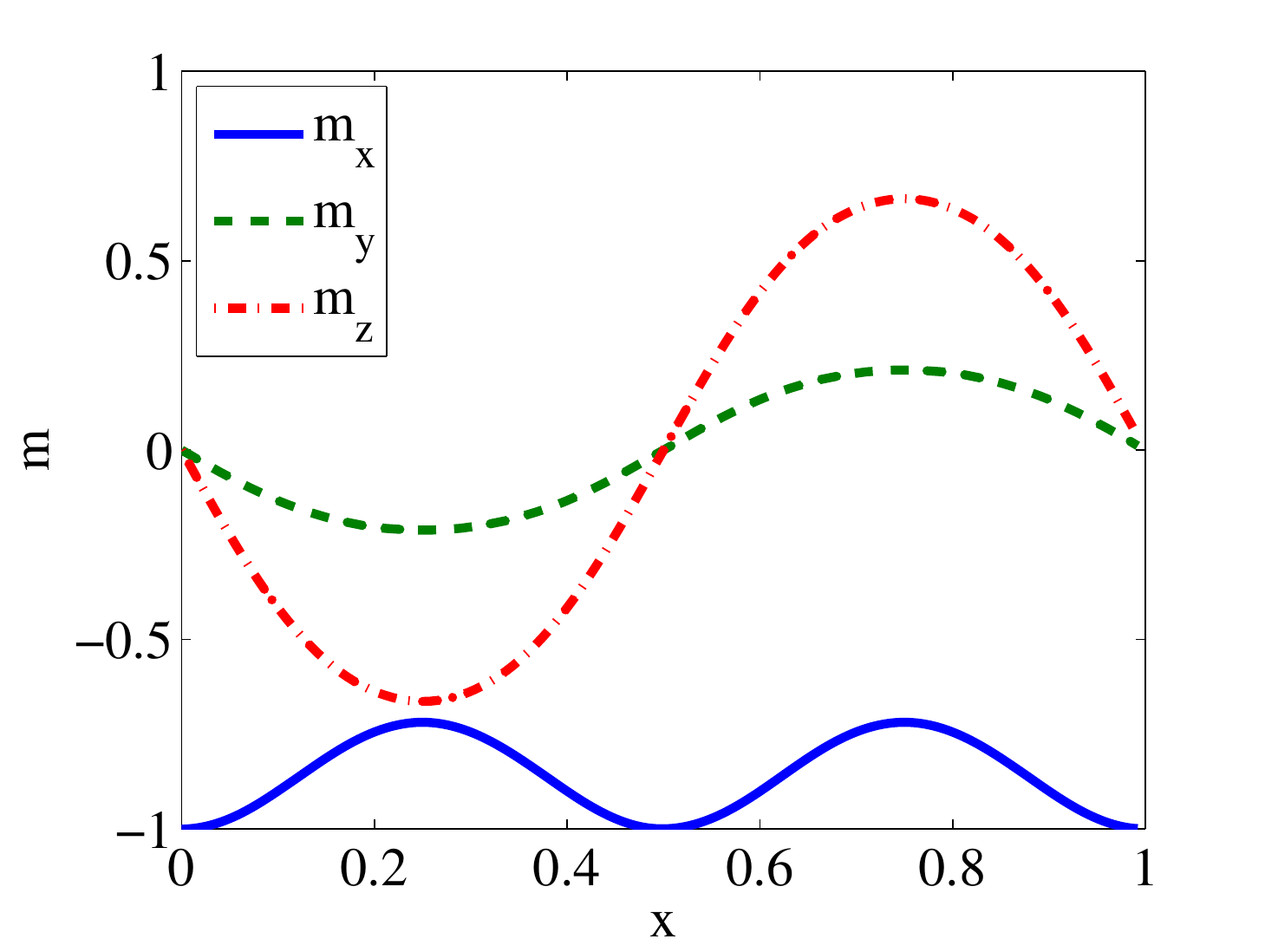}}
}
\subfigure[]{
  \scalebox{0.3}[0.3]{\includegraphics*[viewport=0 0 420 330]{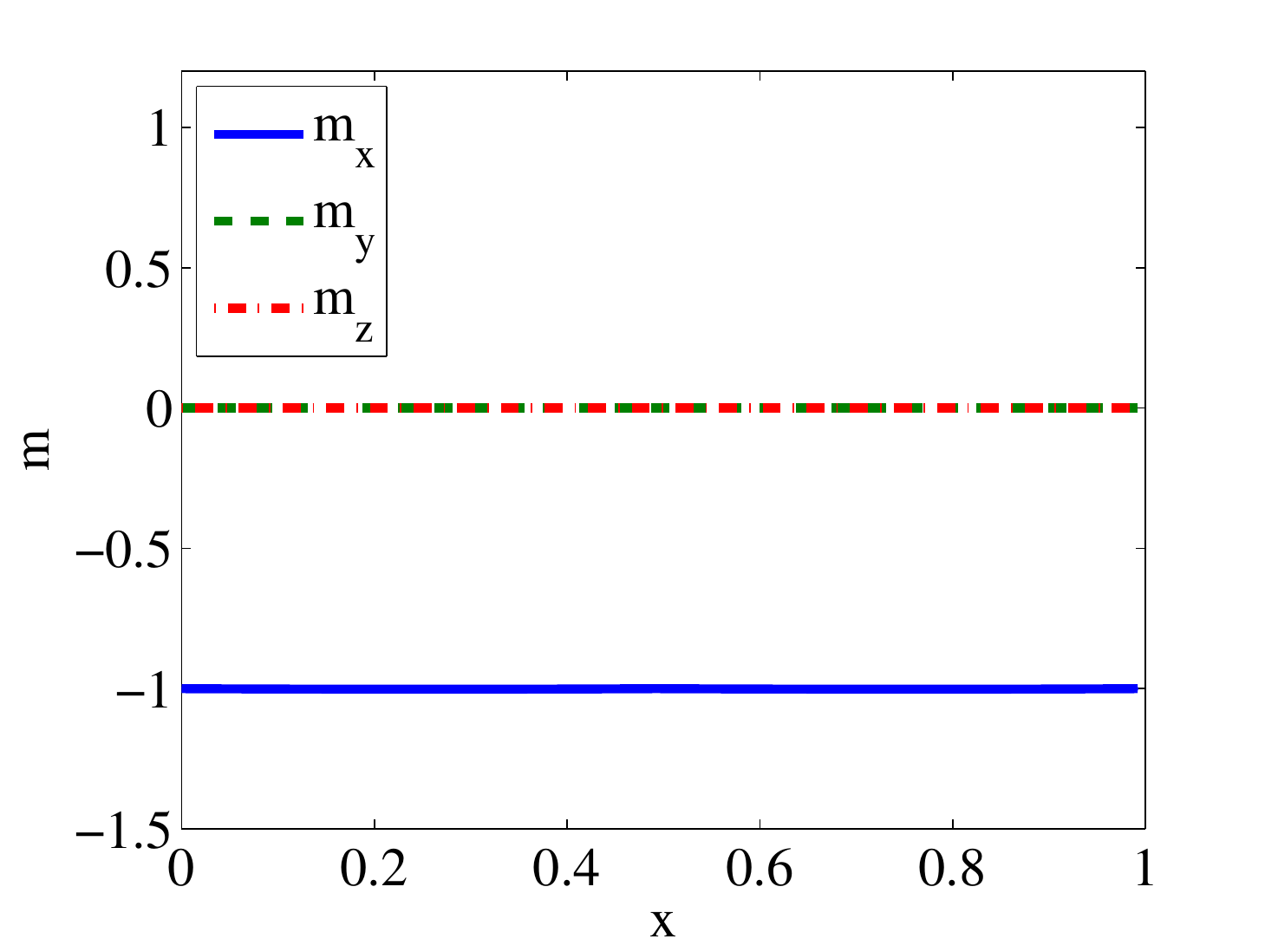}}
}
\subfigure[]{
  \scalebox{0.3}[0.3]{\includegraphics*[viewport=0 0 420 330]{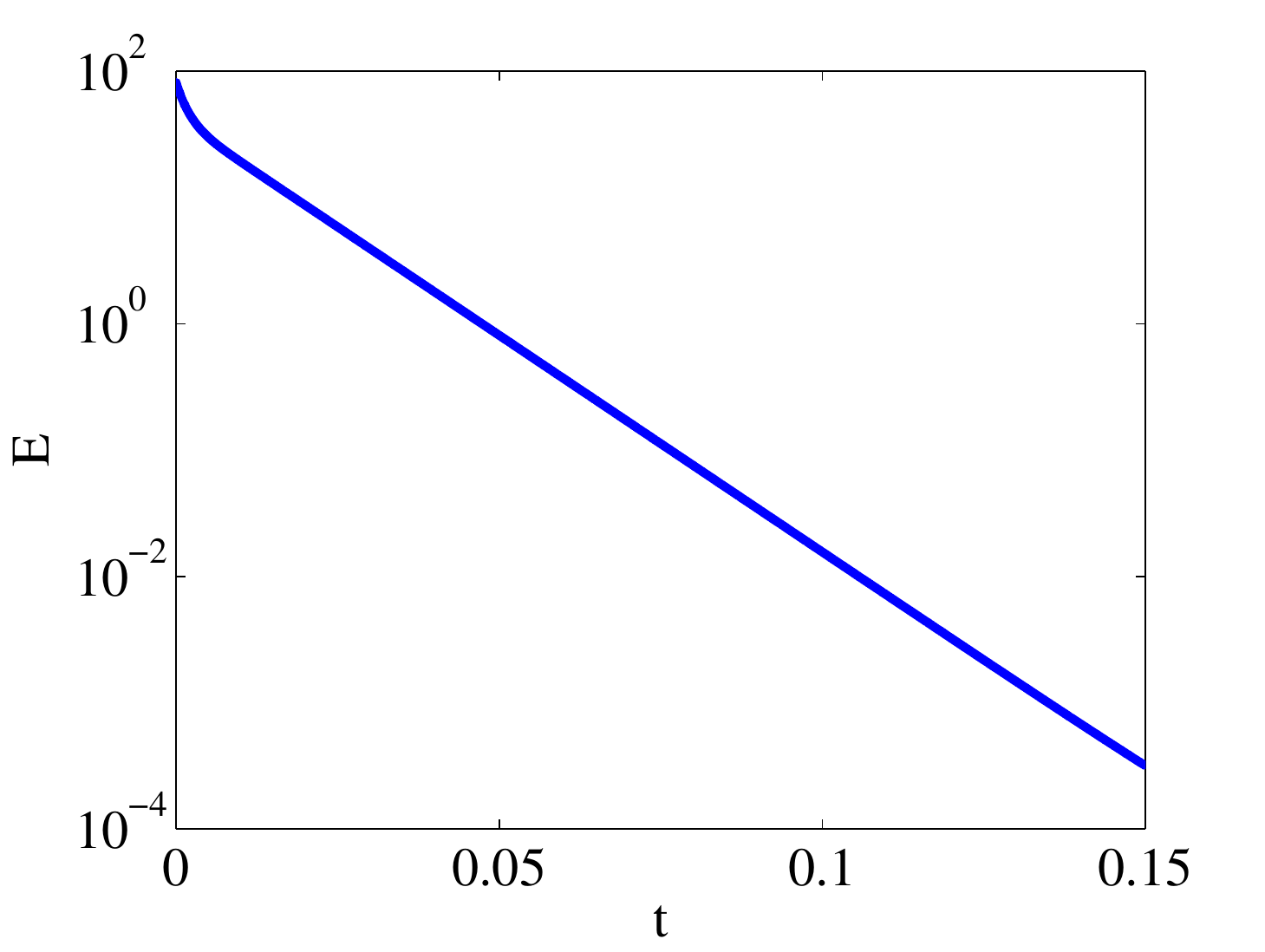}}
}
\caption{(Color online) Numerical simulations of Case~(1), the Landau--Lifshitz--Gilbert
equation in the over-damped limit.  In this case, the magnetization decays
to a constant state.
%
%
%
%
Subfigures (a) and (b) show the magnetization at times $t=0.03$ and $t=0.15$
respectively;
(c) is the energy functional, which exhibits exponential decay after some
transience.  The final orientation is $\left(\phi,\theta\right)=\left(\pi,\pi/2\right)$.}
\label{fig:alpha_LL} 
\end{figure}

\emph{Case 1: Numerical simulations of Eq.~\eqref{eq:gilbert}.}
Equation~\eqref{eq:gilbert} is usually solved by explicit or implicit finite
differences~\cite{Weinan2000}.  We solve the equation by these methods, and
by the explicit spectral method~\cite{Zhu_numerics}.  The accuracy and computational
cost
is roughly the same
in each case, and for simplicity, we therefore employ explicit finite differences;
 it is this method we use throughout the paper.
 Given the initial conditions~\eqref{eq:initial}, each component of the magnetization
 $\Mag=\left(m_x,m_y,m_z\right)$ tends to a constant, the energy
\[
E = \tfrac{1}{2}\int_\Omega{dx}\left|\frac{\partial\Mag}{\partial{x}}\right|^2
\]
decays with time, and $\left|\Mag\right|^2$ retains its initial value $|\Mag|^2=1$.
 After some transience, the decay of the energy functional becomes exponential
 in time.  These results are shown in Fig.~\ref{fig:alpha_LL}

\begin{figure}[htb]
\subfigure[]{
  \scalebox{0.3}[0.3]{\includegraphics*[viewport=0 0 420 330]{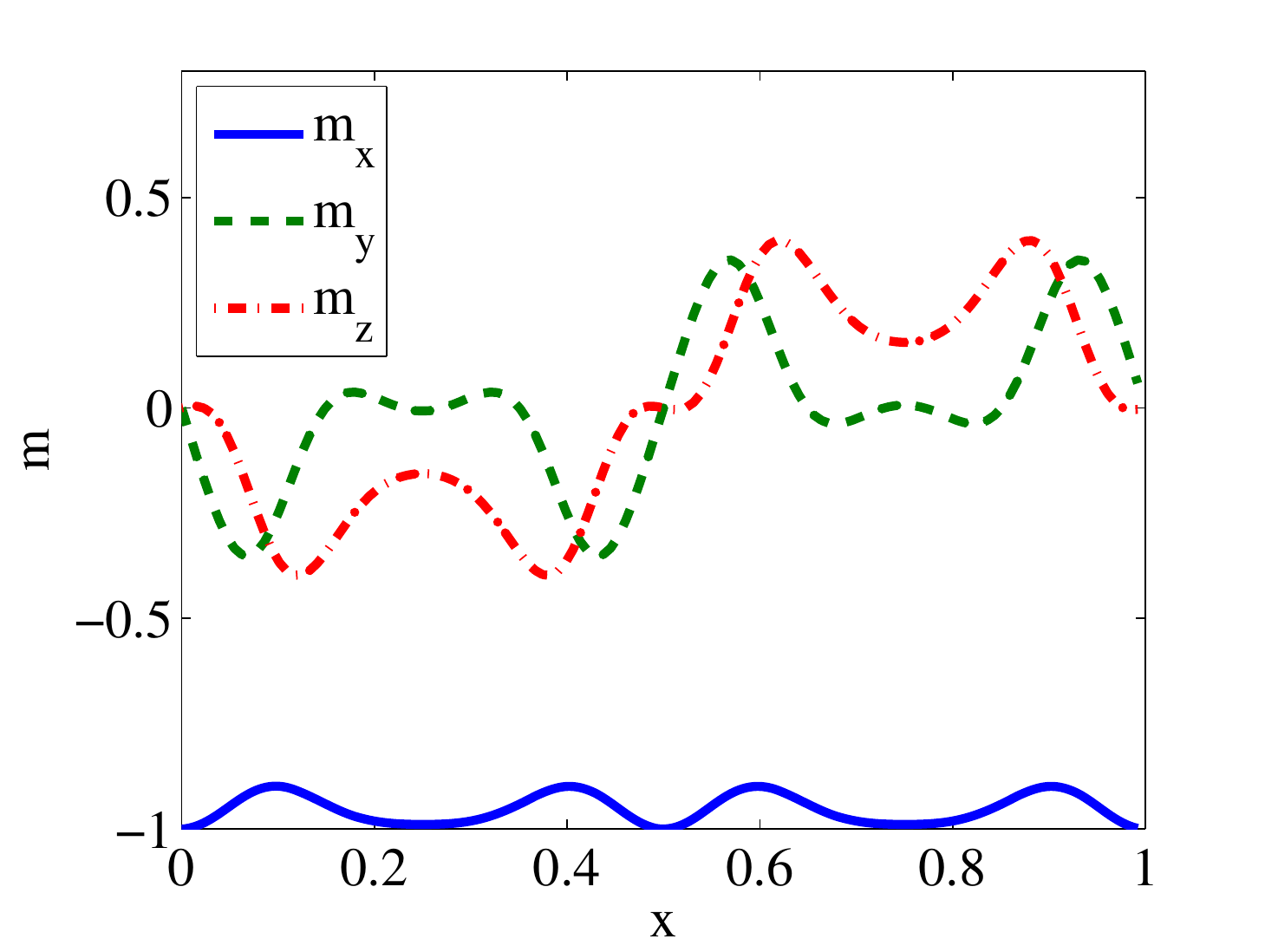}}
}
\subfigure[]{
  \scalebox{0.3}[0.3]{\includegraphics*[viewport=0 0 420 330]{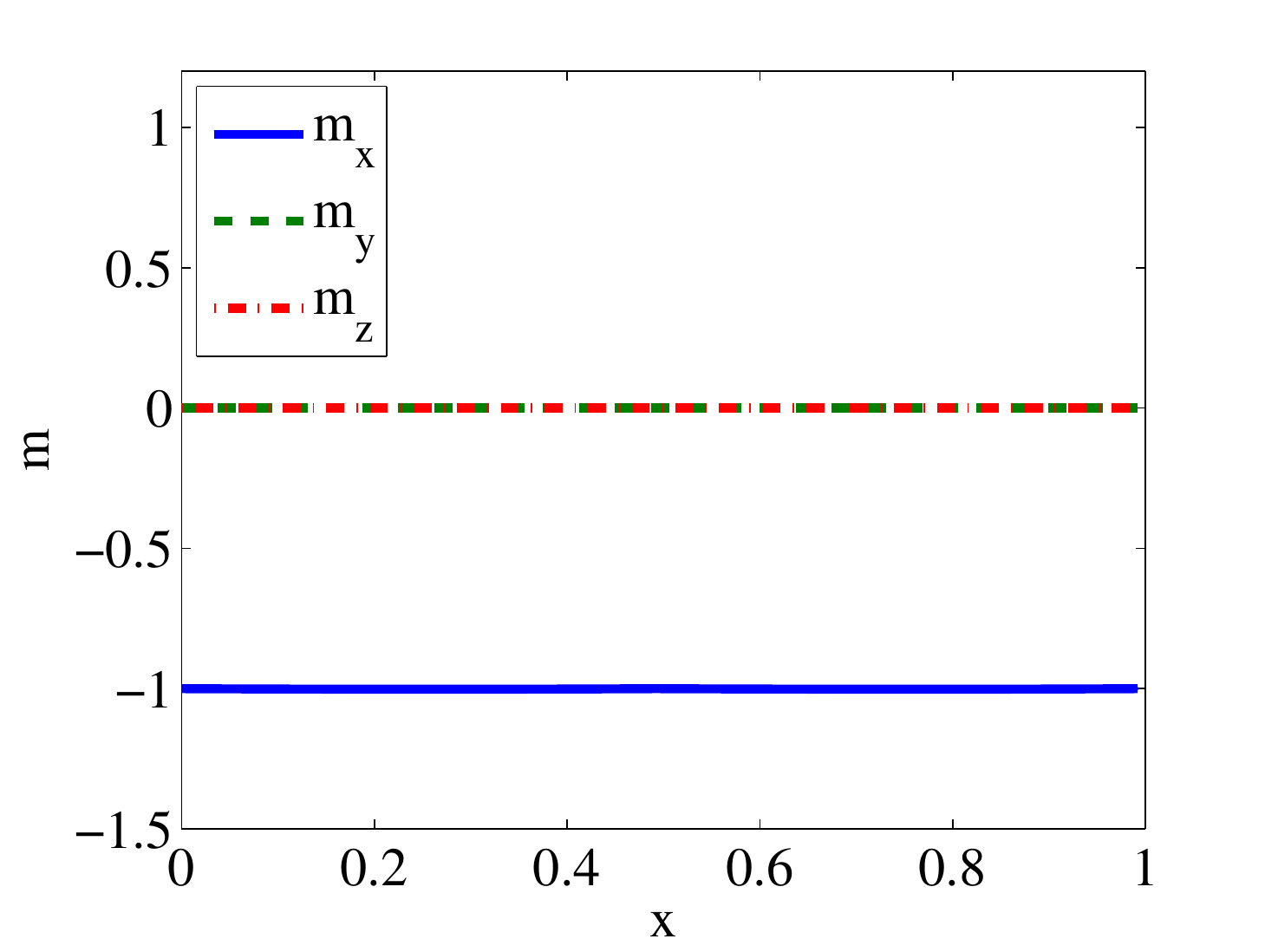}}
}
\subfigure[]{
  \scalebox{0.3}[0.3]{\includegraphics*[viewport=0 0 420 330]{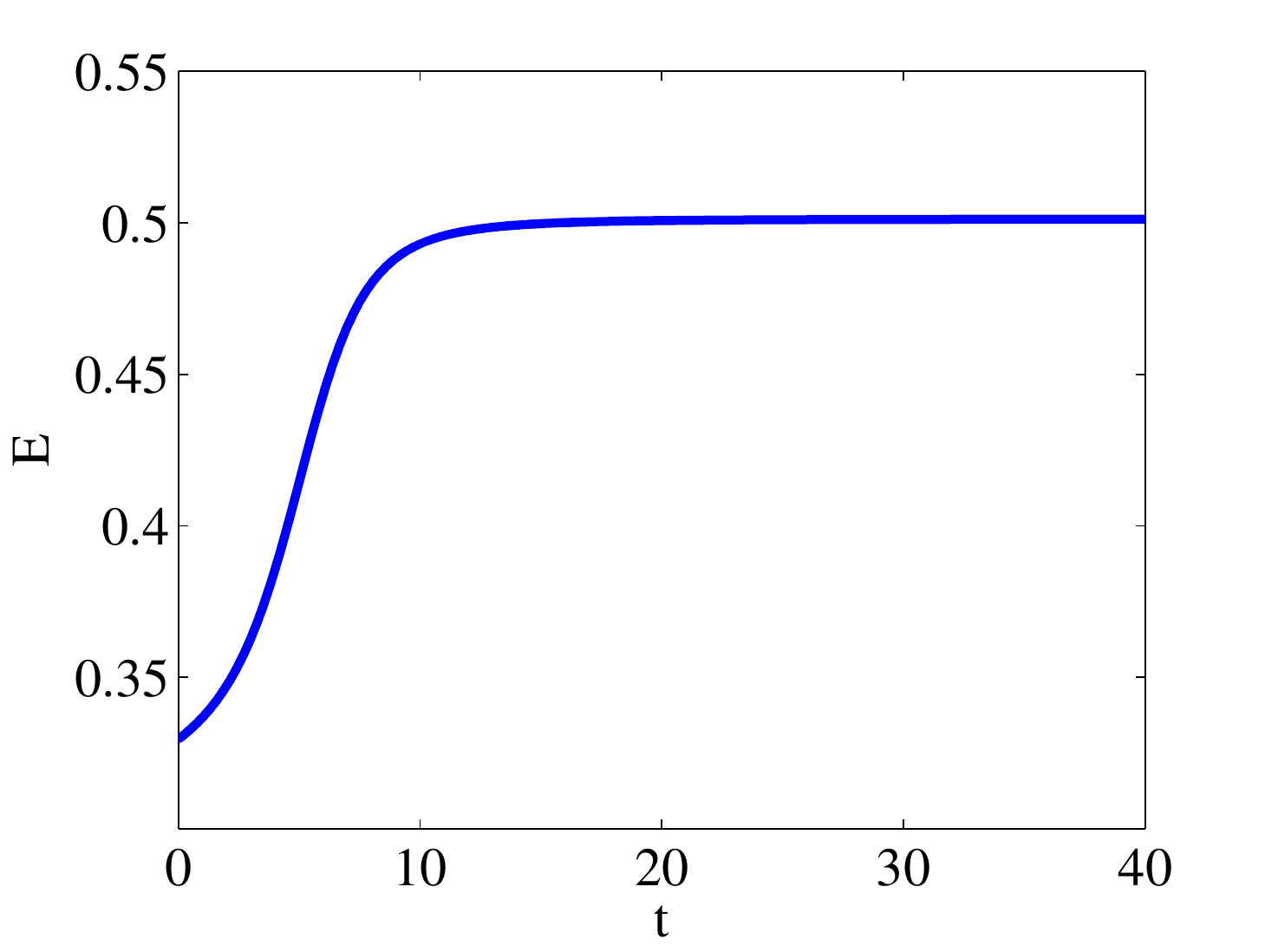}}
}
\caption{(Color online) Numerical simulations of Case~(2), the non-local
Gilbert equation
with with $\alpha<\beta$.  In this case, the energy increases to a constant
value, and the magnetization becomes constant.
%
%
%
%
Subfigures (a) and (b) show the magnetization
at times $t=8$ and $t=40$; (c) is the energy functional.  The final orientation
is $\left(\phi,\theta\right)=\left(\pi,\pi/2\right)$.}
\label{fig:alpha_small} \end{figure}

\emph{Case 2: Numerical simulations of Eq.~\eqref{eq:mag_eqn} with
$\alpha<\beta$}.  Given the smooth initial data~\eqref{eq:initial},
in time each component of the magnetization $\Mag=\left(m_x,m_y,m_z\right)$
decays to zero, while the energy
\[
E=\tfrac{1}{2}\int_\Omega{dx}\Mag\cdot\left(1-\alpha^2\partial_x^2\right)^{-1}\Mag
\]
tends to a constant value.  Given our choice of initial conditions, the energy
in fact \emph{increases} to attain this constant value.  Again the quantity
$\left|\Mag\right|^2$ stays constant.  These results are shown in Fig.~\ref{fig:alpha_small}.
We find similar results when we set $\alpha=0$.

\begin{figure}[htb]
\subfigure[]{
  \scalebox{0.3}[0.3]{\includegraphics*[viewport=0 0 420 330]{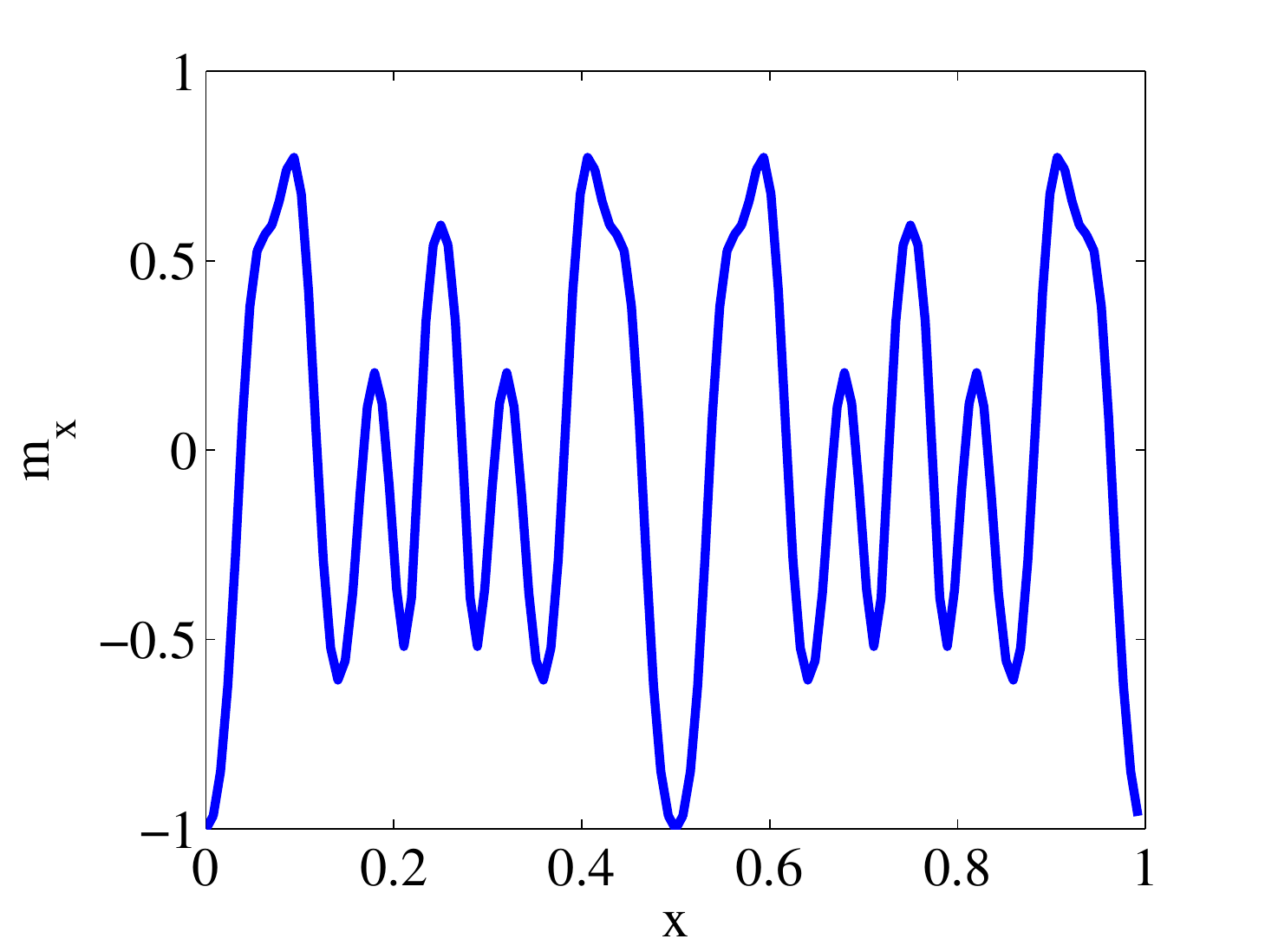}}
}
\subfigure[]{
  \scalebox{0.3}[0.3]{\includegraphics*[viewport=0 0 420 330]{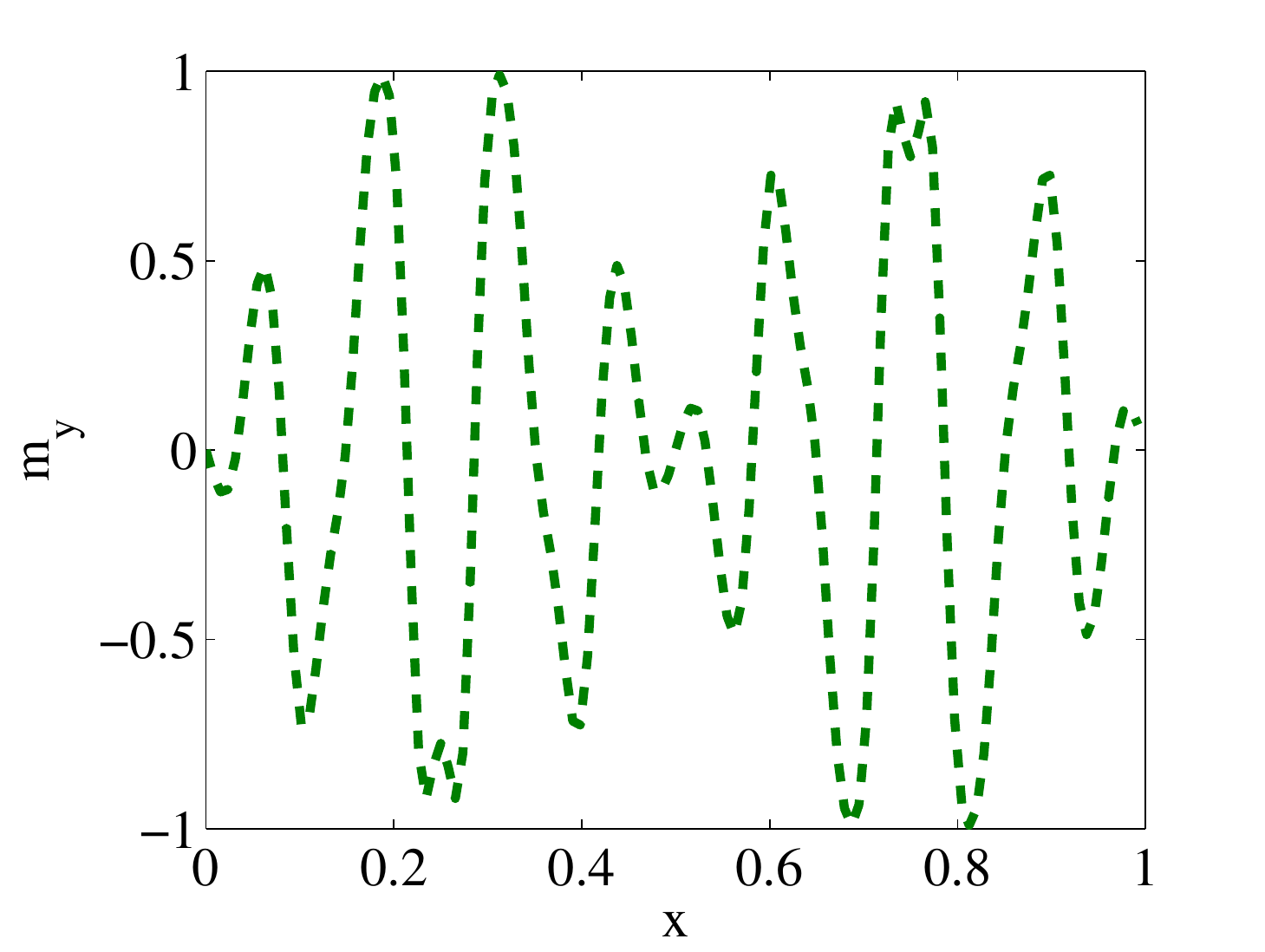}}
}
\subfigure[]{
  \scalebox{0.3}[0.3]{\includegraphics*[viewport=0 0 420 330]{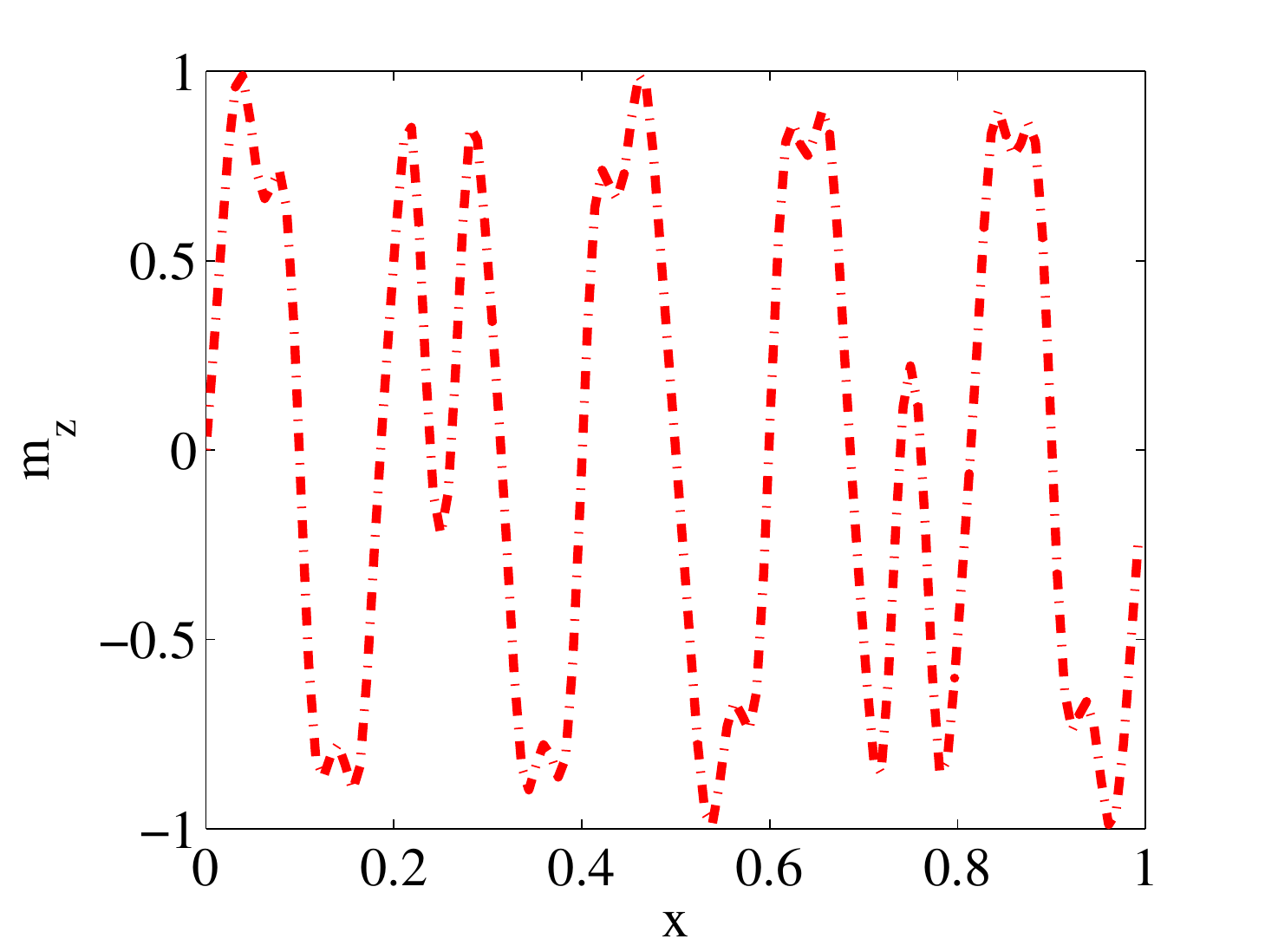}}
}
\subfigure[]{
  \scalebox{0.3}[0.3]{\includegraphics*[viewport=0 0 420 330]{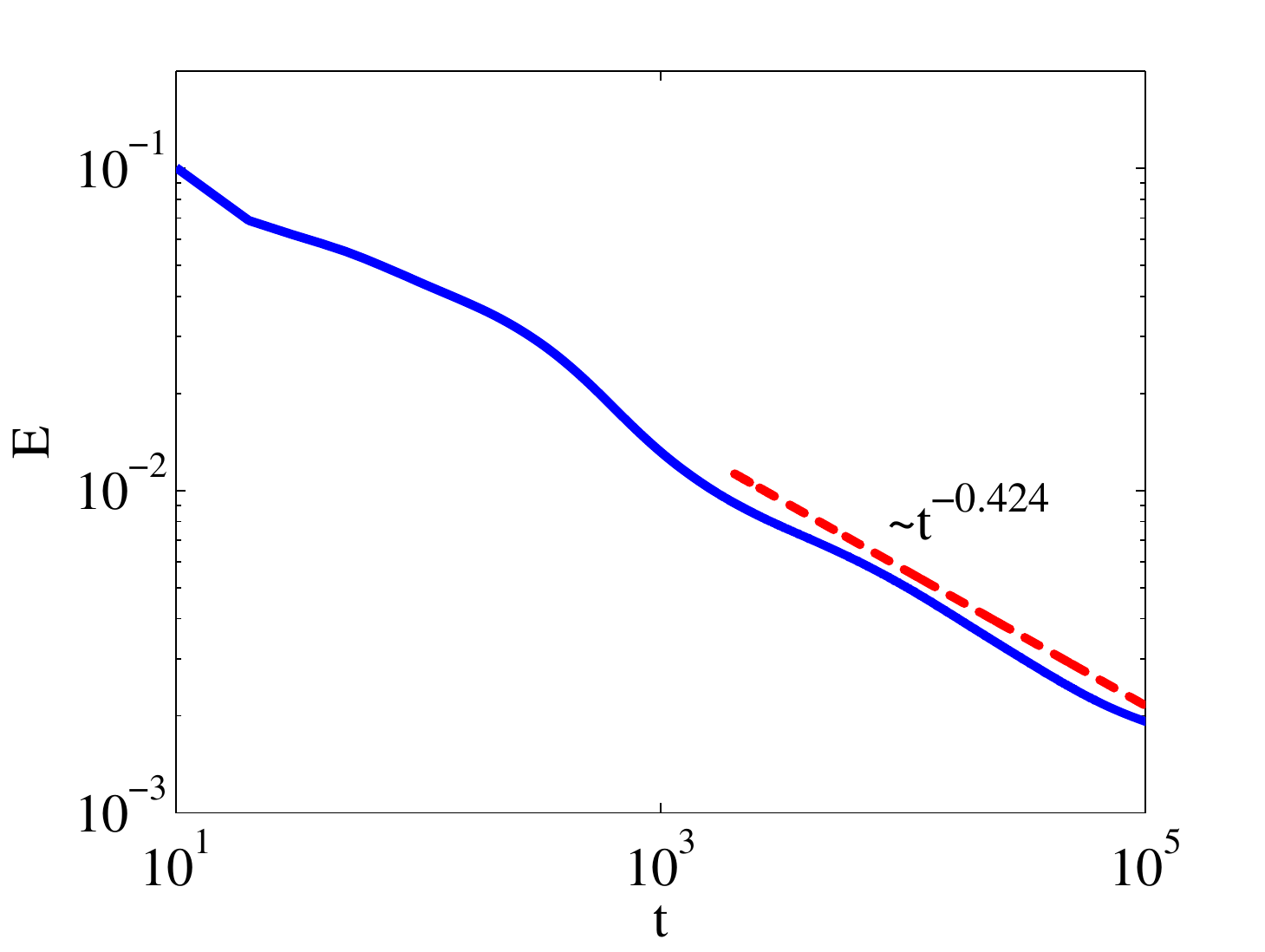}}
}
\subfigure[]{
  \scalebox{0.3}[0.3]{\includegraphics*[viewport=0 0 420 330]{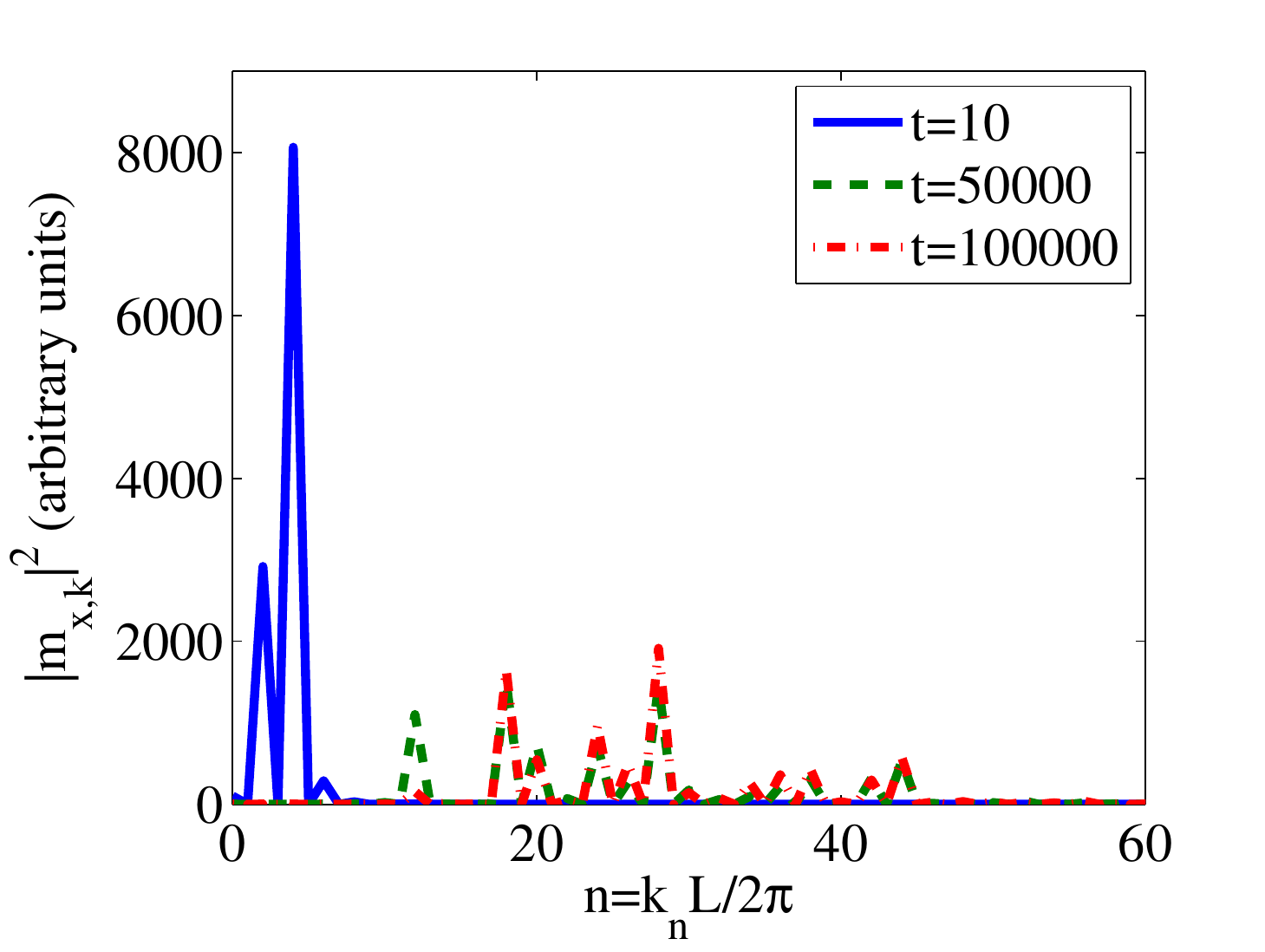}}
}
\caption{(Color online) Numerical simulations of Case~(3), the non-local
Gilbert equation
with $\alpha>\beta$.  In this case, the energy decreases indefinitely, and
the magnetization vector develops finer and finer scales.
%
%
%
%
Subfigures (a), (b), and (c) show the magnetization at time $t=10000$; (d)
is the energy functional, which decreases in time as a power law at late
times.  Subfigure (e) shows the power spectrum of $m_x$;
the integer index $n$ labels the spatial scales: if $k_n$ is a wavenumber,
then the corresponding integer label is $n=k_nL/2\pi$.
}
\label{fig:alpha_big} 
\end{figure}

\emph{Case 3: Numerical simulations of Eq.~\eqref{eq:mag_eqn} with
$\alpha>\beta$}.  Given the smooth initial data~\eqref{eq:initial}, in time
each component of the magnetization $\Mag=\left(m_x,m_y,m_z\right)$ develops
finer and finer scales.  
The development of small scales is driven by the decreasing nature of the
energy functional, which decreases as power law at late times, and is reflected
in snapshots of the power spectrum of the magnetization vector, shown in
Fig.~\ref{fig:alpha_big}.  As the system evolves, there
\begin{table}[h!b!p!]
\begin{tabular}{|c|c|c|c|c|}
\hline
Case&Length scales&Energy&Outcome as $t\rightarrow\infty$&Linear Stability\\
\hline
(1)&$\beta=0$, $\delta{E}/\delta{\Mag}=-\partial_x^2\Mag$&Decreasing&Constant
state&Stable\\
(2)&$\alpha<\beta$&Increasing&Constant state&Stable\\
(3)&$\alpha>\beta$&Decreasing&Development of finer and finer scales&Unstable\\
\hline
\end{tabular}
\caption{Summary of the forms of Eq.~\eqref{eq:mag_eqn} studied.}
\label{tab:table_summary}
\end{table}
is a transfer
of large amplitudes to higher wave numbers.  This transfer slows down at
late
times, suggesting that the rate at which the solution roughens tends to zero,
as $t\rightarrow\infty$.
The evolution preserves the symmetry of the magnetization
vector $\Mag\left(x,t\right)$ under parity transformations.  This is
seen by comparing Figs.~\ref{fig:initial} and~\ref{fig:alpha_big}.
The energy is a decaying function of time, while the quantity
$\left|\Mag\right|^2$ stays constant.  We find similar results for the case
when $\beta=0$.

These results can be explained qualitatively as follows.  In Case~(1), the
energy functional exacts a penalty for the formation of gradients.  The energy
decreases with time and the the system evolves into a state in
which no magnetization gradients are present, that is, a constant state.
On the other hand, we have demonstrated that in Case~(2), when $\alpha<\beta$,
the energy increases to a constant value.  Since in the non-local model, the
energy functional represents the cost of forming smooth spatial structures,
an increase in energy produces a smoother magnetization field, a process
that continues until the  magnetization reaches a constant value.  
Finally, in Case~(3), when $\alpha>\beta$, the energy functional
decreases, and this decrease corresponds to a roughening of the magnetization
field, as seen in Fig.~\ref{fig:alpha_big}. 
In Sec.~\ref{sec:mag_dens} we
shall show that Case~(2) is stable to small perturbations around a constant
state, while Case~(3) is unstable.  Furthermore, we note that Case~(2) and
Case~(3) differ only by a minus sign in Eq.~\eqref{eq:mag_eqn}, and are therefore
related by time reversal.
These results are summarized in Table~\ref{tab:table_summary}.

The solutions of Eqs.~\eqref{eq:mag_eqn} and~\eqref{eq:gilbert} do not become
singular.  This is not surprising: the manifest
conservation of $\left|\Mag\right|^2$ in Eqs.~\eqref{eq:mag_eqn}
and~\eqref{eq:gilbert} provides a pointwise bound on the magnitude of the
solution, preventing blow-up.  Any addition
to Eq.~\eqref{eq:mag_eqn} that breaks this conservation law gives rise to
the possibility of singular solutions, and it is to this possility that we
now turn.

\section{Coupled density-magnetization equations}
\label{sec:mag_dens}
In this section we study a coupled density-magnetization equation
pair that admit singular solutions.
 We investigate the linear stability
of the equations and examine the conditions for instability.  We find that
the stability or otherwise of a constant state is controlled by the magnetization
and density values of that state, and by the relative magnitude of the problem
length scales.  Using numerical and analytical techniques, we investigate
the emergence and self-interaction of singular solutions.

The equations we study are as follows,
\begin{subequations}
\begin{equation}
\frac{\partial\rho}{\partial t}=\frac{\partial}{\partial{x}}\left[\rho\left(\mu_\rho\frac{\partial}{\partial{x}}\frac{\delta{E}}{\delta\rho}+\bm{\mu}_{m}\cdot\frac{\partial}{\partial{x}}\frac{\delta{E}}{\delta\bm{m}}\right)\right],
\end{equation}
\begin{equation}
\frac{\partial\Mag}{\partial t} = \frac{\partial}{\partial x}\left[\Mag\left(\mu_\rho\frac{\partial}{\partial{x}}\frac{\delta{E}}{\delta\rho}+\bm{\mu}_{m}\cdot\frac{\partial}{\partial{x}}\frac{\delta{E}}{\delta\Mag}\right)\right]+\Mag\times\left(\bm{\mu}_{\Mag}\times\frac{\delta{E}}{\delta\Mag}\right),
\end{equation}%
\label{eq:mag_dens}%
\end{subequations}%
where we set
\[
\mu_\rho=1,\qquad\frac{\partial{E}}{\partial\rho}=-\left(1-\alpha_\rho^2\partial_x^2\right)^{-1}\rho,
\]
and, as before,
\[
\bm{\mu}_{m} = \left(1-\beta_\Subm^2\partial_x^2\right)^{-1}\Mag,\qquad
\frac{\delta E}{\delta\Mag} = \left(1-\alpha_\Subm^2\partial_x^2\right)^{-1}\Mag.
\]
These equations have been introduced by Holm, Putkaradze and Tronci in~\cite{Darryl_eqn1},
using a kinetic-theory description.
%
%
The density and the magnetization vector are driven by the velocity
\begin{equation}
V=\mu_\rho\frac{\partial}{\partial{x}}\frac{\delta{E}}{\delta\rho}+\bm{\mu}_{m}\cdot\frac{\partial}{\partial{x}}\frac{\delta{E}}{\delta\bm{m}}.
\label{eq:velocity}
\end{equation}
The velocity advects the ratio $|\Mag|/\rho$ by
\[
\left(\frac{\partial}{\partial{t}}-V\frac{\partial}{\partial{x}}\right)\frac{|\Mag|}{\rho}=0.
\]
We have the system energy
\begin{equation}
E = \tfrac{1}{2}\int_\Omega{dx}\Mag\cdot\left(1-\alpha_\Subm^2\partial_x^2\right)^{-1}\Mag
- \tfrac{1}{2}\int_\Omega{dx}\rho\left(1-\alpha_\rho^2\partial_x^2\right)^{-1}\rho,
\label{eq:energy_mag_dens}
\end{equation}
and, given a non-negative density, the second term is always non-positive.
 This represents an energy of attraction, and we therefore expect singularities
 in the magnetization vector to arise from a collapse of the particle density
 due to the ever-decreasing energy of attraction.  There are three length
 scales in the problem that control the time evolution: the
 ranges $\alpha_\Subm$ and $\alpha_\rho$ of the potentials in Eq.~\eqref{eq:energy_mag_dens},
 and the smoothening length $\beta_\Subm$.
\subsection*{Linear stability analysis}
We study the linear stability of the constant state $\left(\Mag,\rho\right)=\left(\Mag_0,\rho_0\right)$.
We evaluate the smoothened values of this constant solution as follows,
\begin{eqnarray*}
\left(1-\alpha_\rho^2\partial_x^2\right)^{-1}\rho_0&=&f\left(x\right),\\
\rho_0&=&f\left(x\right)-\alpha_\rho^2\frac{d^2 f}{dx^2},\\
f\left(x\right)&=&\rho_0+A\sinh\left(x/\alpha_\rho\right)+B\cosh\left(x/\alpha_\rho\right).
\end{eqnarray*}
For periodic or infinite boundary conditions, the constants $A$ and
$B$ are in fact zero and thus 
\begin{equation}
\left(1-\alpha_\rho\partial_x^2\right)^{-1}\rho_0=\rho_0,
\label{eq:smooth_const}
\end{equation}
and similarly $\bm{\mu}_0=\left(\delta E/\delta\Mag\right)_{\Mag_0}=\Mag_0$.
 The result~\eqref{eq:smooth_const} guarantees that the constant state $\left(\Mag_0,\rho_0\right)$
 is indeed a solution of Eq.~\eqref{eq:mag_dens}.
 
We study a solution $\left(\Mag,\rho\right)=\left(\Mag_0+\delta\Mag,\rho_0+\delta\rho\right)$,
which represents a perturbation away from the constant state.  By assuming
that $\delta\Mag$ and $\delta\rho$ are initially small in magnitude, we obtain
the following linearized equations for the perturbation density and magnetization,
\begin{subequations}
\begin{equation*}
\frac{\partial}{\partial t}\delta\rho=-\rho_0\frac{\partial^2}{\partial{x}^2}\left(1-\alpha_\rho^2\partial_x^2\right)^{-1}\delta\rho+\rho_0\frac{\partial^2}{\partial{x}^2}\left(1-\alpha_\Subm^2\partial_x^2\right)^{-1}\Mag_0\cdot\delta\Mag,
\end{equation*}
\begin{multline*}
\frac{\partial}{\partial t}\delta\Mag =\Mag_0\left[-\frac{\partial^2}{\partial{x}^2}\left(1-\alpha_\rho^2\partial_x^2\right)^{-1}\delta\rho+\frac{\partial^2}{\partial{x}^2}\left(1-\alpha_\Subm^2\partial_x^2\right)^{-1}\Mag_0\cdot\delta\Mag\right]\\
+\Mag_0\times\Big\{\Mag_0\times\left[\left(1-\alpha_m^2\partial_x^2\right)^{-1}\delta\Mag-\left(1-\beta_m^2\partial_x^2\right)^{-1}\delta\Mag\right]\Big\}.
\end{multline*}%
\label{eq:mag_dens_linear}%
\end{subequations}%
 For $\Mag_0\neq0$ we may choose two unit vectors $\hat{\bm{n}}_1$ and $\hat{\bm{n}}_2$
 such that $\Mag_0/|\Mag_0|$, $\hat{\bm{n}}_1$ and $\hat{\bm{n}}_2$ form
 an orthonormal triad (that is, we have effected a change of basis).  We
 then study the quantities $\delta\rho$, $\delta\chi$, $\delta\xi_1$ and
 $\delta\xi_2$, where
\[
\delta\chi=\Mag_0\cdot\delta\Mag,\qquad\delta\xi_1=\hat{\bm{n}}_1\cdot\delta\Mag,\qquad\delta\xi_2=\hat{\bm{n}}_2\cdot\delta\Mag.
\]
We obtain the linear equations
\begin{equation*}
\frac{\partial}{\partial t}\delta\rho=-\rho_0\frac{\partial^2}{\partial{x}^2}\left(1-\alpha_\rho^2\partial_x^2\right)^{-1}\delta\rho+\rho_0\frac{\partial^2}{\partial{x}^2}\left(1-\alpha_\Subm^2\partial_x^2\right)^{-1}\delta\chi,
\end{equation*}
\begin{equation*}
\frac{\partial}{\partial t}\delta\chi =-|\Mag_0|^2\frac{\partial^2}{\partial{x}^2}\left(1-\alpha_\rho^2\partial_x^2\right)^{-1}\delta\rho+|\Mag_0|^2\frac{\partial^2}{\partial{x}^2}\left(1-\alpha_\Subm^2\partial_x^2\right)^{-1}\delta\chi,
\end{equation*}
\begin{equation*}
\frac{\partial}{\partial t}\delta\xi_i=\left[\left(1-\beta_\Subm^2\partial_x^2\right)^{-1}-\left(1-\alpha_\Subm^2\partial_x^2\right)^{-1}\right]\delta\xi_i,\qquad
i=1,2.
\end{equation*}%
By focusing on a single-mode disturbance with wave number $k$ we obtain the
following system of equations
\begin{equation*}
\frac{d}{dt}\left(\begin{array}{c}\delta\rho \\ \delta\chi \\ \delta\xi_1
\\ \delta\xi_2\end{array}\right)=
\left(\begin{array}{cccc}
\frac{\rho_0k^2}{1+\alpha_\rho^2k^2}&-\frac{\rho_0k^2}{1+\alpha_\Subm^2k^2}&0&0\\
\frac{|\Mag_0|^2k^2}{1+\alpha_\rho^2k^2}&-\frac{|\Mag_0|^2k^2}{1+\alpha_\Subm^2k^2}&0&0\\
0&0&\frac{1}{1+\beta_\Subm^2k^2}-\frac{1}{1+\alpha_\Subm^2k^2}&0\\
0&0&0&\frac{1}{1+\beta_\Subm^2k^2}-\frac{1}{1+\alpha_\Subm^2k^2}
\end{array}\right)
\left(\begin{array}{c}\delta\rho \\ \delta\chi \\ \delta\xi_1
\\ \delta\xi_2\end{array}\right),
\end{equation*}
%
%
%
%
%
%
with eigenvalues
\begin{equation}
\sigma_0=0,\qquad 
\sigma_1=\frac{\rho_0k^2}{1+\alpha_\rho^2k^2}-\frac{|\Mag_0|^2k^2}{1+\alpha_\Subm^2k^2},\qquad
\sigma_2=\frac{1}{1+\beta_\Subm^2k^2}-\frac{1}{1+\alpha_\Subm^2k^2}.
\end{equation}
The eigenvalues are the growth rate of the disturbance $\left(\delta\rho,\delta\chi,\delta\xi_1,\delta\xi_2\right)$~\cite{ChandraFluids}.
 There are two routes to instability, when $\sigma_1>0$, or when $\sigma_2>0$.
  The first route leads to an instability when
\[
\sigma_1>0,\qquad \frac{\rho_0}{|\Mag_0|^2}>\frac{1+\alpha_\rho^2k^2}{1+\alpha_\Subm^2k^2},
\]
while the second route leads to instability when
\[
\sigma_2>0,\qquad \alpha_\Subm>\beta_\Subm.
\] 
%
%
%
%
%
%
%
 
\begin{figure}[htb]
\subfigure[]{
  \scalebox{0.45}[0.45]{\includegraphics*[viewport=0 0 420 330]{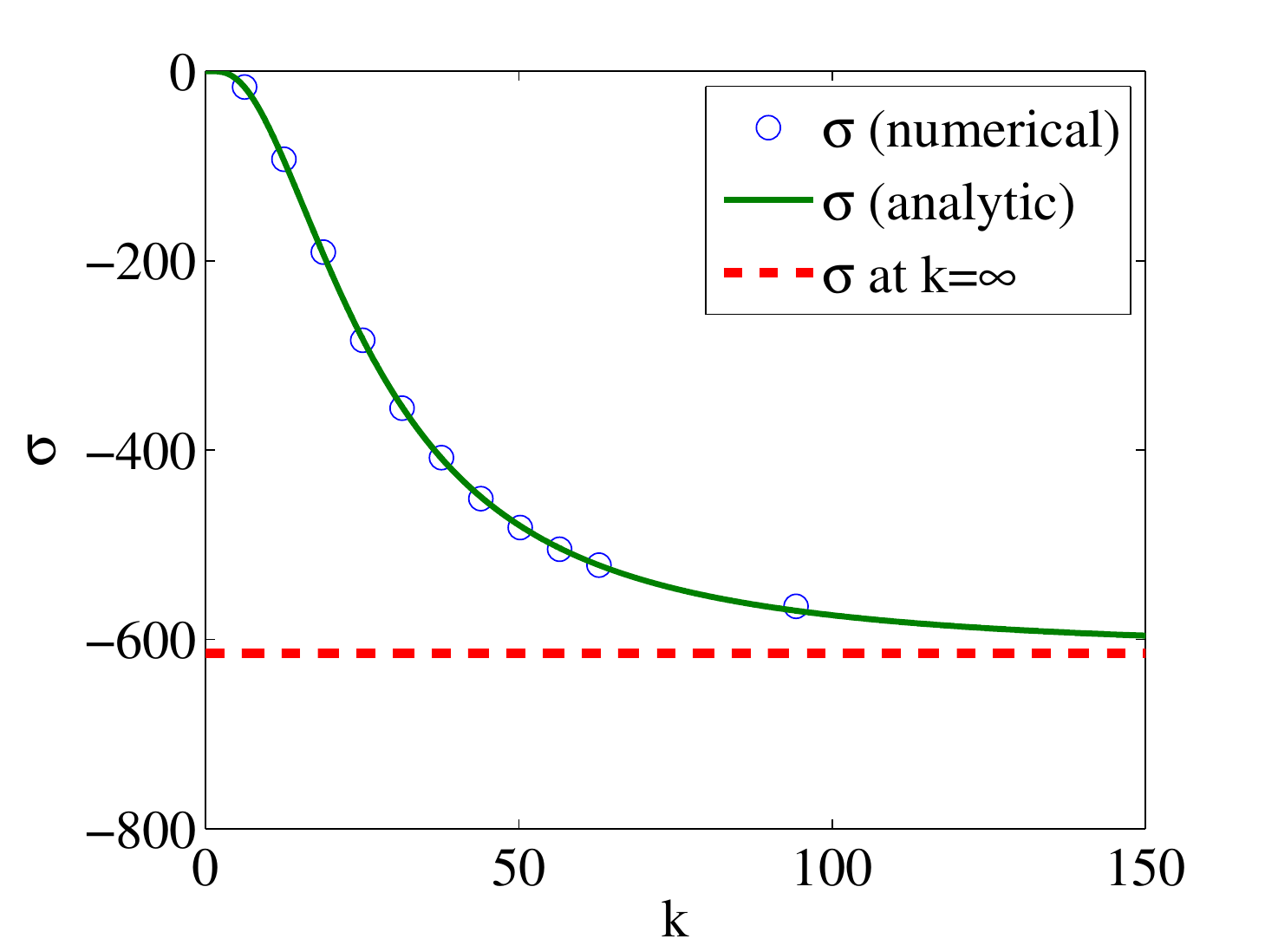}}
}
\subfigure[]{
  \scalebox{0.45}[0.45]{\includegraphics*[viewport=0 0 420 330]{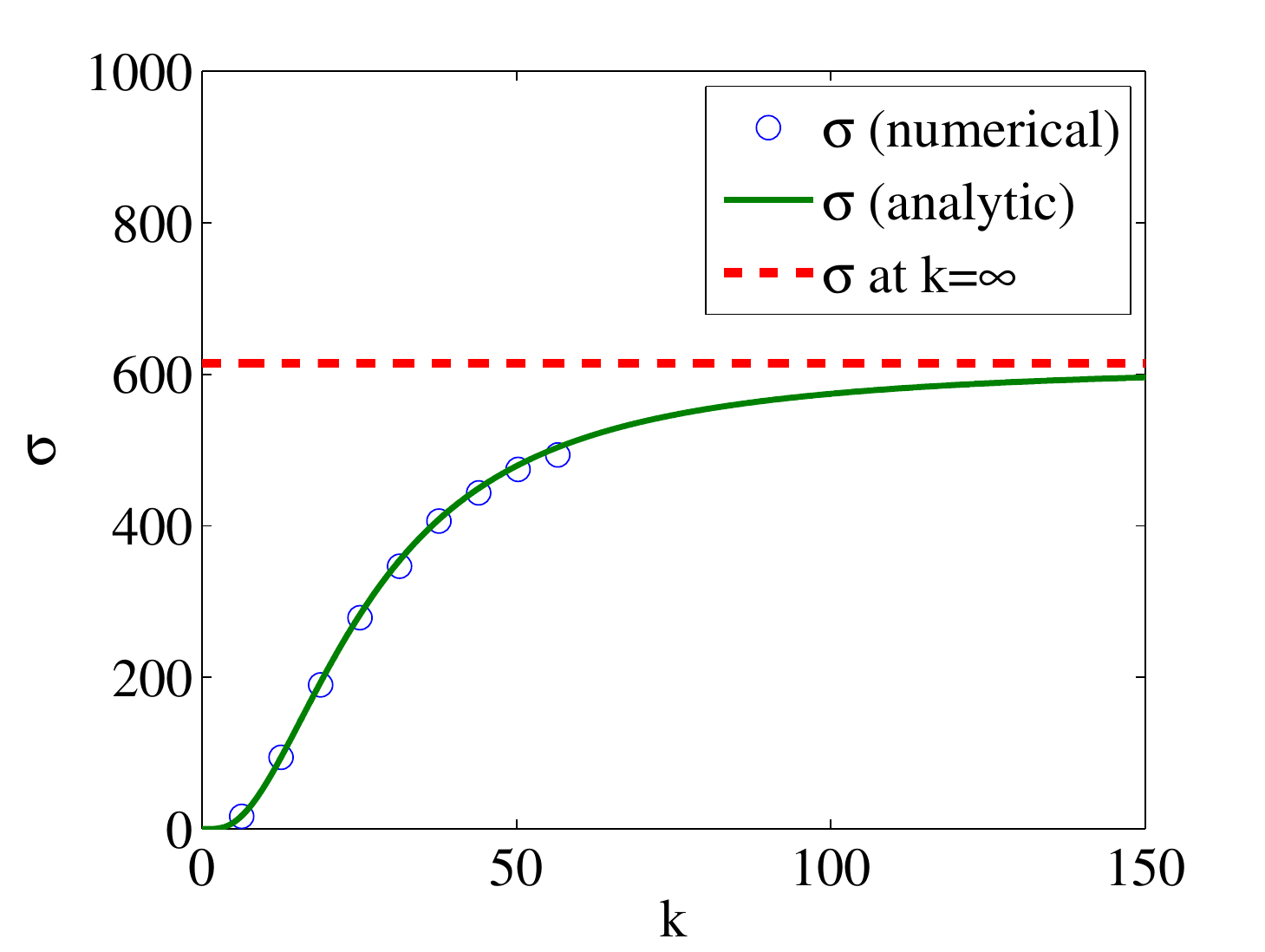}}
}
\caption{(Color online) The first route to instability.   Subfigure (a) shows
the growth
rate $\sigma_1$ for $\alpha_\Subm<\beta_\Subm<\alpha_\rho$, with negativity
indicating a stable equilibrium; (b) gives the growth rate $\sigma_1$
for $\alpha_\rho<\alpha_\Subm<\beta_\Subm$, with positivity indicating an
unstable equilibrium.  We have set $|\Mag_0|=\rho_0=1$.
}
\label{fig:growth_rates} 
\end{figure}

\begin{figure}[htb]
\subfigure[]{
  \scalebox{0.45}[0.45]{\includegraphics*[viewport=0 0 420 330]{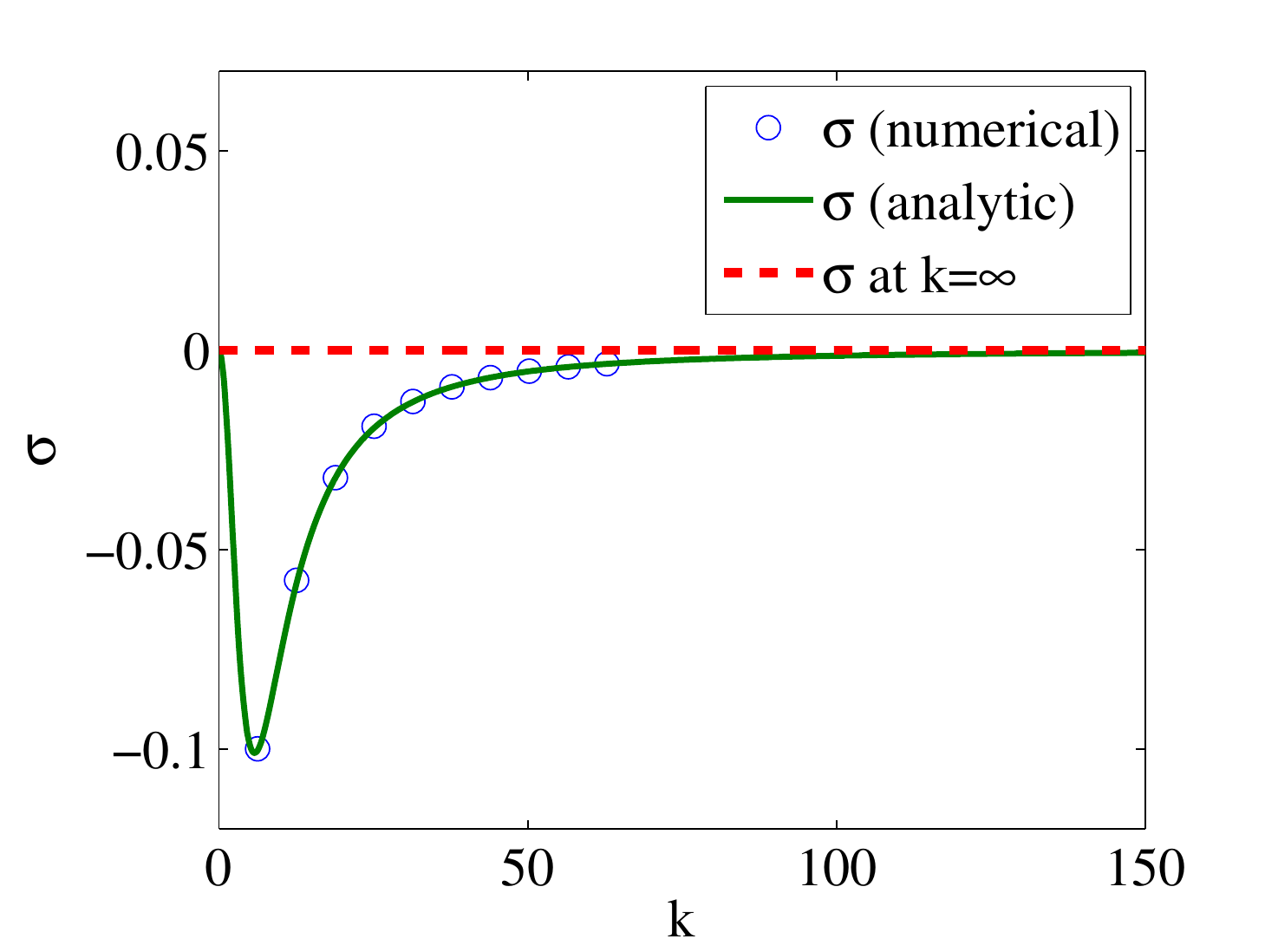}}
}
\subfigure[]{
  \scalebox{0.45}[0.45]{\includegraphics*[viewport=0 0 420 330]{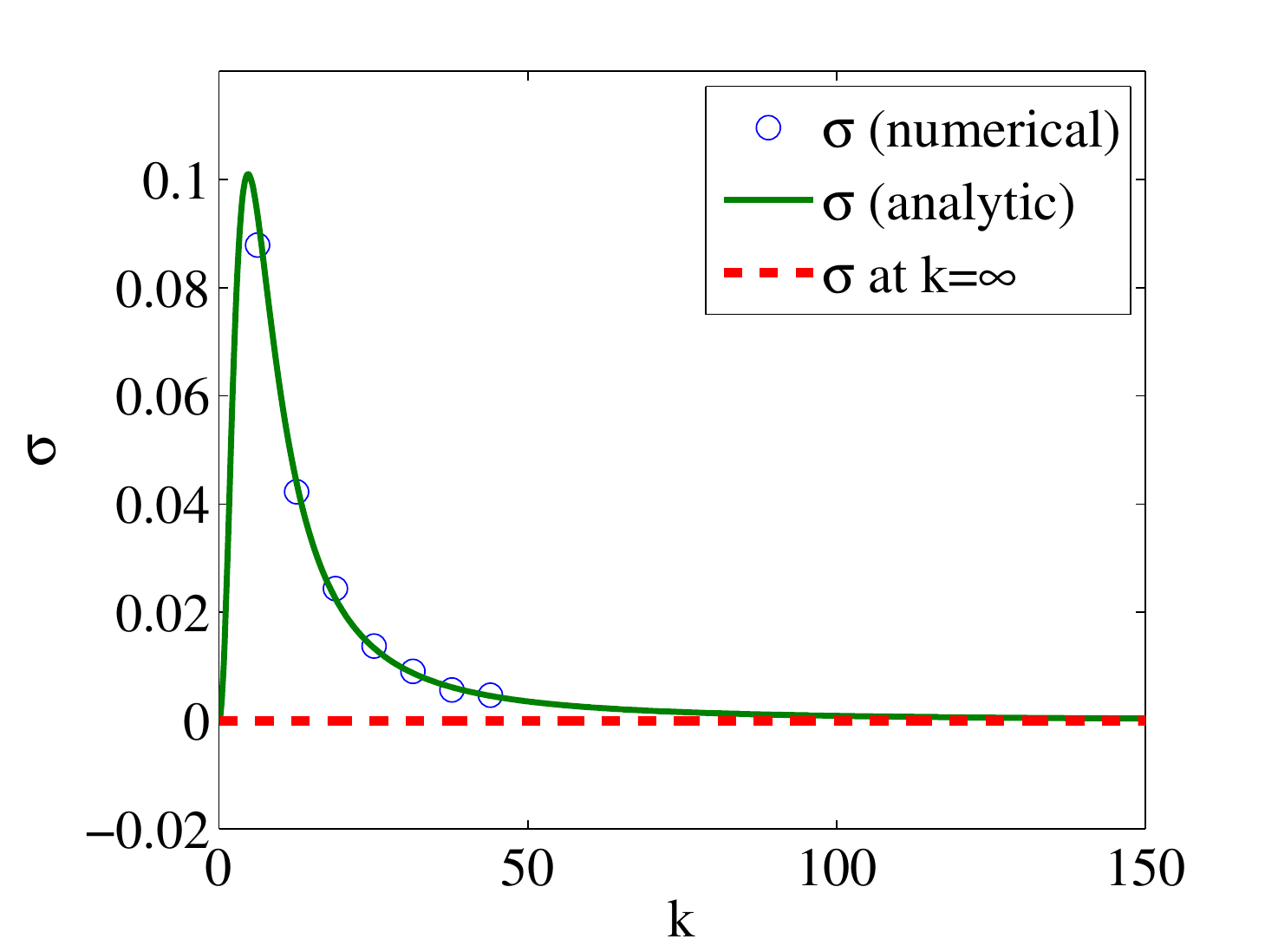}}
}
\caption{(Color online) The second route to instability.   Subfigure (a)
shows the growth
rate $\sigma_2$ for $\alpha_\Subm<\beta_\Subm$, with negativity
indicating a stable equilibrium; (b) gives the growth rate $\sigma_2$
for $\alpha_\Subm>\beta_\Subm$, with positivity indicating an unstable equilibrium.
 We have set $|\Mag_0|=\rho_0=1$.   
}
\label{fig:growth_rates2} 
\end{figure}

We have plotted the growth rates for the case when $\rho_0=|\Mag_0|^2=1$,
and compared the theory with numerical simulations.  There is excellent agreement
at low wave numbers, although the numerical simulations become less accurate
at high wave numbers.  This can be remedied by increasing the resolution
of the simulations.  These plots are shown in Figs.~\ref{fig:growth_rates}
and~\ref{fig:growth_rates2}.
The growth rates $\sigma_{1,2}$ are parabolic in $k$ at small $k$; $\sigma_1$
saturates at large $k$, while $\sigma_2$ attains a maximum and decays at
large $k$.  The growth rates can be positive or negative, depending on the
initial configuration,
and on the relationship between the problem length scales.
 In contrast to some standard instabilities of pattern formation (e.g. Cahn--Hilliard~\cite{Argentina2005}
 or Swift--Hohenberg~\cite{OjalvoBook}), the $\sigma_1$-unstable state
 becomes more unstable at higher wave numbers (smaller scales), thus preventing
 the `freezing-out' of the instability by a reduction of the box size~\cite{Argentina2005}.
  The growth at small scales is limited,
however, by the saturation in $\sigma$ as $k\rightarrow\infty$.  Heuristically,
 this can be explained as follows: at higher wave number, the disturbance
  $\left(\delta\rho,\delta\chi,\delta\xi_1,\delta\xi_2\right)$ gives rise
  to more and more peaks per unit length.  This makes merging events increasingly
  likely, so that peaks combine to form larger peaks, enhancing the growth
  of the disturbance.

Recall in Sec.~\ref{sec:gilbert} that the different behaviors of the magnetization
equation~\eqref{eq:mag_eqn} are the result of a competition between the length
scales
$\alpha_\Subm$ and $\beta_\Subm$.  For $\alpha_\Subm<\beta_\Subm$ the initial
(large-amplitude) disturbance tends to a constant, while for $\alpha_\Subm>\beta_\Subm$
the initial disturbance develops finer and finer scales.  In this section,
we have shown that the coupled density-magnetization equations are linearly
stable when $\alpha_\Subm<\beta_\Subm$, while the reverse case is unstable.
 In contrast to the first route to instability, the growth rate $\sigma_2$,
 if positive, admits a maximum.  This is obtained by setting $\sigma_2'\left(k\right)=0$.
  Then the maximum growth rate occurs at a scale
\begin{equation*}
\lambda_{\mathrm{max}}:=2\pi k_{\mathrm{max}}^{-1}=2\pi\sqrt{\alpha_\Subm\beta_\Subm}.
\end{equation*}
Thus, the scale at which the disturbance is most unstable is determined
by the geometric mean of $\alpha_\Subm$ and $\beta_\Subm$.  Given a disturbance
$\left(\delta\rho,\delta\chi,\delta\xi_1,\delta\xi_2\right)$ with a range
of modes initially present, the instability selects the disturbance on the
scale $\lambda_{\mathrm{max}}$.  This disturbance develops a large amplitude
and a singular solution subsequently emerges.  It is to this aspect of the
problem that we now turn.

\subsection*{Singular solutions}
%
%
%
%
 In this section we show that a finite weighted sum of delta functions satisfies
 the partial differential equations~\eqref{eq:mag_dens}.  Each delta function
 has the
 interpretation of a particle or clumpon, whose weights and positions satisfy
 a finite set of ordinary differential equations.  We investigate the two-clumpon
 case analytically and show that the clumpons tend to a state in which they
 merge, diverge, or are separated by a fixed distance.  In each case, we
 determine the final state of the clumpon magnetization.
 

To verify that singular solutions are possible, let us substitute the ansatz
\begin{equation}
\rho\left(x,t\right)=\sum_{i=1}^M a_i\left(t\right)\delta\left(x-x_i\left(t\right)\right),\qquad
\Mag\left(x,t\right)=\sum_{i=1}^M \bm{b}_i\left(t\right)\delta\left(x-x_i\left(t\right)\right),
\label{eq:clumpon_sln}
\end{equation}
into the weak form of equations~\eqref{eq:mag_dens}.  Here we sum over the
different components of the singular solution (which we call clumpons).
In this section we work on the infinite domain $x\in\left(-\infty,\infty\right)$.
 The weak form of the equations is obtained by testing Eqs.~\eqref{eq:mag_dens}
with once-differentiable functions $\phi\left(x\right)$ and $\bm{\psi}\left(x\right)$,
\begin{subequations}
\begin{equation}
\frac{d}{dt}\Int{dx}\rho\left(x,t\right)\phi\left(x\right)=-\Int{dx}\phi'\left(x,t\right)\left(\mu_\rho\frac{\partial}{\partial{x}}\frac{\delta{E}}{\delta\rho}+\bm{\mu}_{\bm{m}}\cdot\frac{\partial}{\partial{x}}\frac{\delta{E}}{\delta\bm{m}}\right),
\end{equation}
\begin{multline}
\frac{d}{dt}\Int{dx}\Mag\left(x,t\right)\cdot\bm{\psi}\left(x\right) 
= -\Int{dx}\bm{\psi}'\left(x\right)\cdot\Mag\left(x,t\right)\left(\mu_\rho\frac{\partial}{\partial{x}}\frac{\delta{E}}{\delta\rho}
+\bm{\mu}_{\Mag}\cdot\frac{\partial}{\partial{x}}\frac{\delta{E}}{\delta\Mag}\right)
\\
+\Int{dx}\bm{\psi}\left(x\right)\cdot\left[\Mag\times\left(\bm{\mu}_{\Mag}\times\frac{\delta{E}}{\delta\Mag}\right)\right],
\end{multline}%
\label{eq:mag_dens_weak}%
\end{subequations}%
Substitution of the ansatz~\eqref{eq:clumpon_sln} into the weak equations~\eqref{eq:mag_dens_weak}
yields the relations
\begin{equation}
\frac{da_i}{dt}=0,\qquad \frac{dx_i}{dt}=-V\left(x_i\right),\qquad\frac{d\bm{b}_i}{dt}=\bm{b}_i\times\left(\bm{\mu}\times\frac{\delta{E}}{\delta\Mag}\right)\left(x_i\right),\qquad
i\in\{1,...,M\},
\label{eq:clumpon_evolution}
\end{equation}
where $V$ and $\left(\bm{\mu}\times\left({\delta{E}}/{\delta\Mag}\right)\right)$
are obtained from the ansatz~\eqref{eq:clumpon_sln} and are evaluated at
$x_i$.  Note that the density weights $a_i$ and the magnitude of the
weights $\bm{b}_i$ remain constant in time.

We develop further understanding of the clumpon dynamics  by studying the
two-clumpon version of Eqs.~\eqref{eq:clumpon_evolution}.  Since the weights
$a_1$, $a_2$, $|\bm{b}_1|$, and $|\bm{b}_2|$ are constant, two variables
%
%
%
%
suffice to describe the interaction: the relative separation $x=x_1-x_2$
of the clumpons,
and the cosine of the angle between the clumpon magnetizations, $\cos\varphi=\bm{b}_1\cdot\bm{b}_2/|\bm{b}_1||\bm{b}_2|$.
Using the properties
of the kernel $H\left(0\right)=1$, $H'\left(0\right)=0$, we derive the equations
%
%
%
%
%
\begin{subequations}
\begin{equation}
\frac{dx}{dt}=M H_{\alpha_\rho}'\left(x\right)- B_1 H'_{\alpha_m}\left(x\right)H_{\beta_m}\left(x\right)
- B_2 H_{\alpha_m}'\left(x\right)y,\qquad y=\cos\varphi
\end{equation}
\begin{equation}
\frac{dy}{dt}=B_2\left(1-y^2\right)\left[H_{\beta_m}\left(x\right)-H_{\alpha_m}\left(x\right)\right],
\label{eq:xtheta_b}
\end{equation}%
\label{eq:xtheta}%
\end{subequations}%
where $M=a_1+a_2$, $B_1=|\bm{b}_1|^2+|\bm{b}_2|^2$, and $B_2=2|\bm{b}_1||\bm{b}_2|$
are constants.  
Equations~\eqref{eq:xtheta} form a dynamical system whose
properties we now investigate using phase-plane analysis~\cite{StrogatzBook}.
 We note first
\begin{figure}[htb]
\subfigure[]{
  \scalebox{0.4}[0.4]{\includegraphics*[viewport=0 0 400 320]{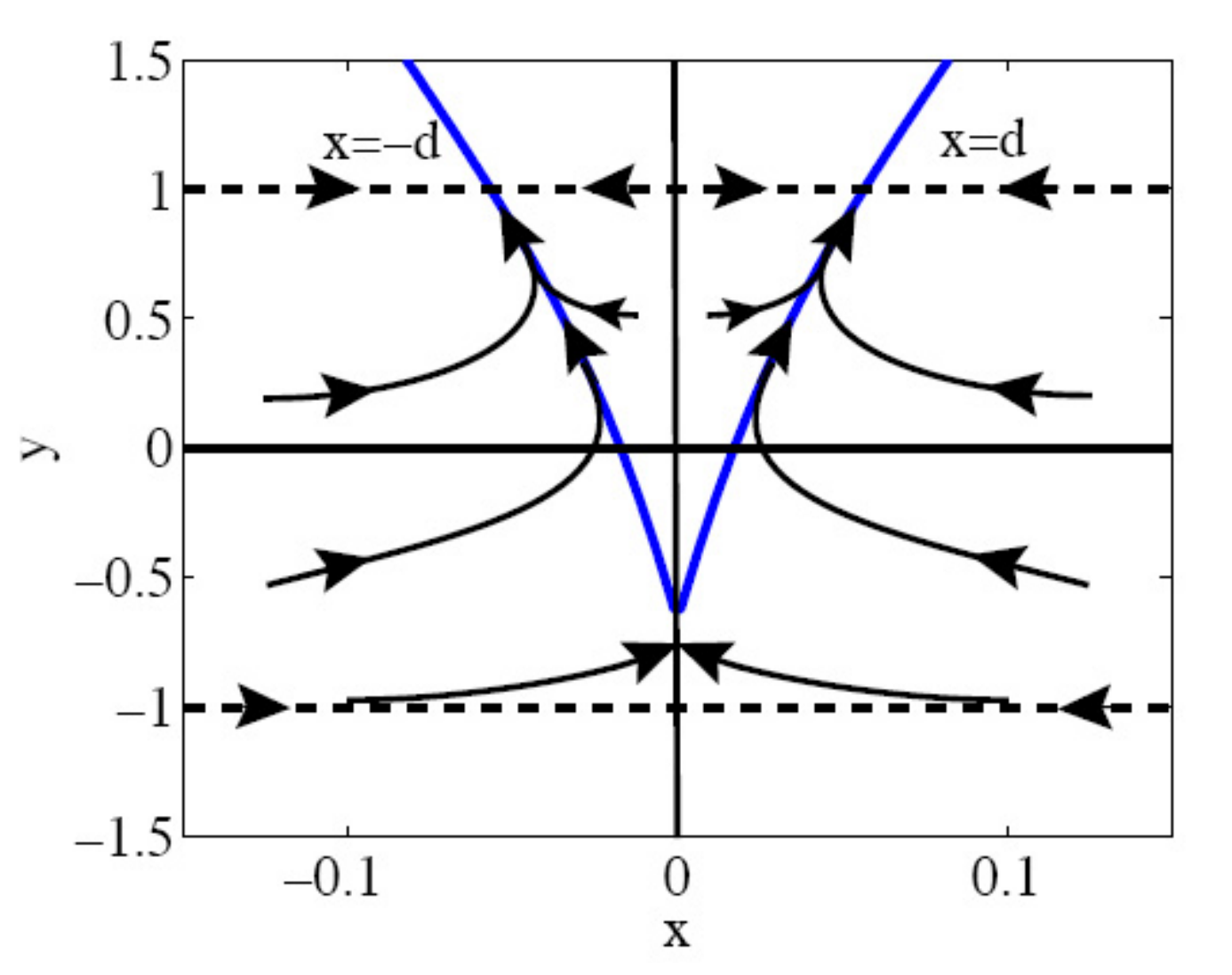}}
}
\subfigure[]{
  \scalebox{0.4}[0.4]{\includegraphics*[viewport=0 0 400 320]{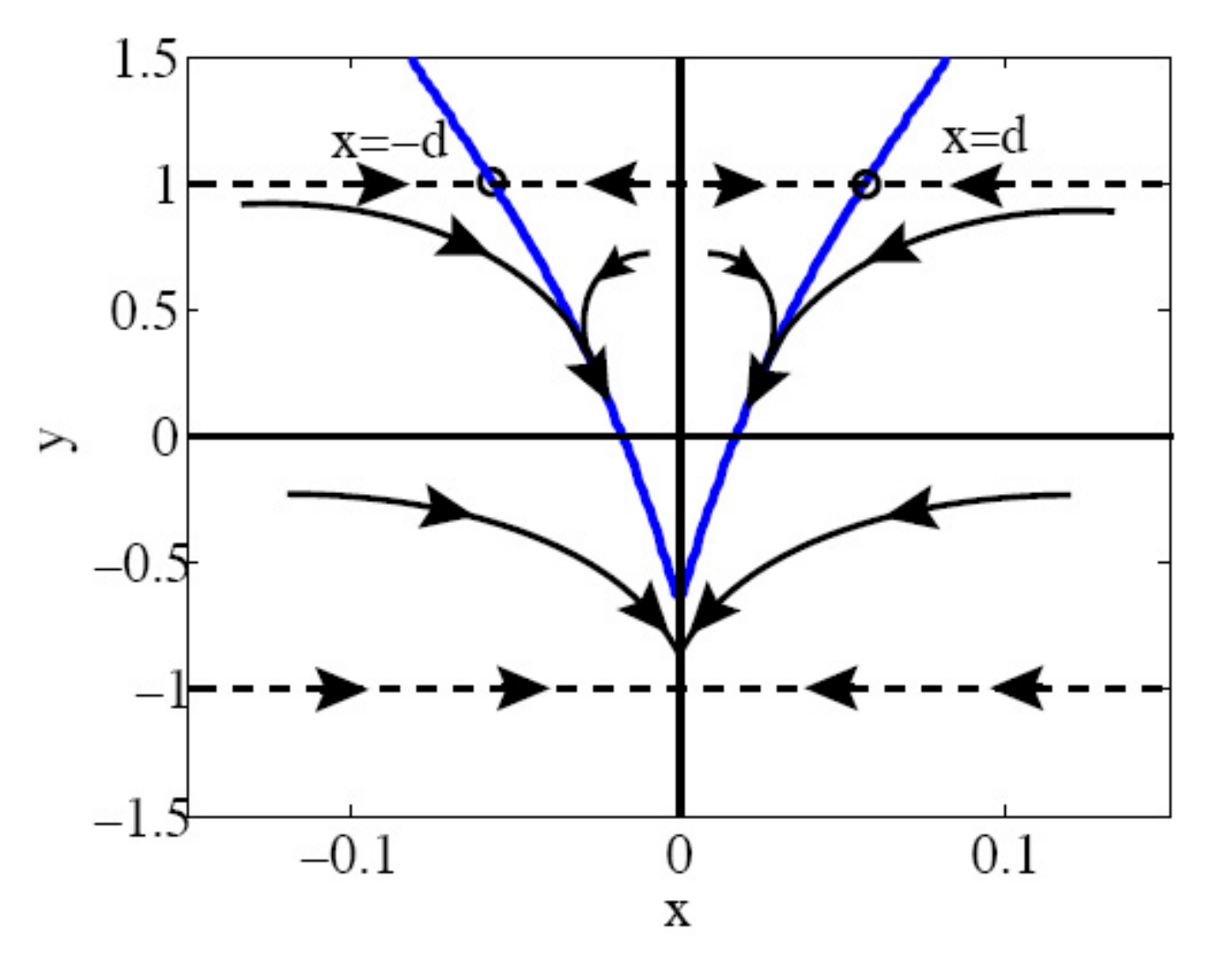}}
}
\caption{(Color online) The nullcline $dx/dt=0$ of the two-clumpon dynamical
system with $\alpha_m<\alpha_\rho$.  The
region contained inside the dotted lines $y=\pm1$ gives the allowed values
of the dynamical variables $\left(x,y\right)$.
 Subfigure~(a)
shows the case when $\beta_m<\alpha_m$.  The stable equilibria of the
system are $\left(x,y\right)=\left(\pm{d},1\right)$ and the line $x=0$.
 All initial conditions flow into one of these equilibrium states; subfigure
 (b) shows the case when $\alpha_m<\beta_m$.  Initial conditions confined
 to the
 line $y=1$ flow into the fixed point $\left(\pm{d},1\right)$,
  while all other initial conditions flow into the line $x=0$.
}
\label{fig:nullclines} 
\end{figure} 
\begin{figure}[htb]
\subfigure[]{
  \scalebox{0.4}[0.4]{\includegraphics*[viewport=0 0 400 320]{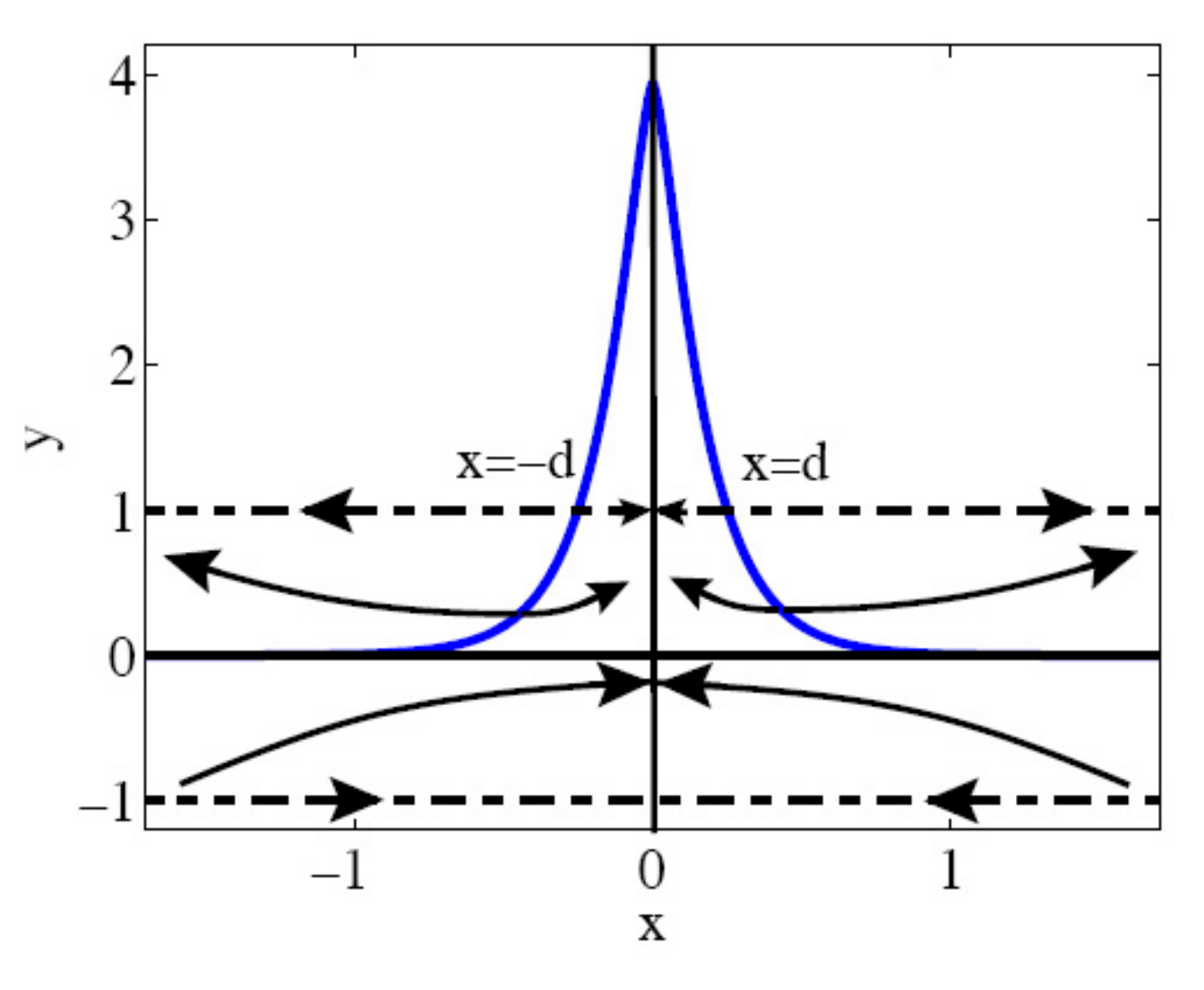}}
}
\subfigure[]{
  \scalebox{0.4}[0.4]{\includegraphics*[viewport=0 0 400 320]{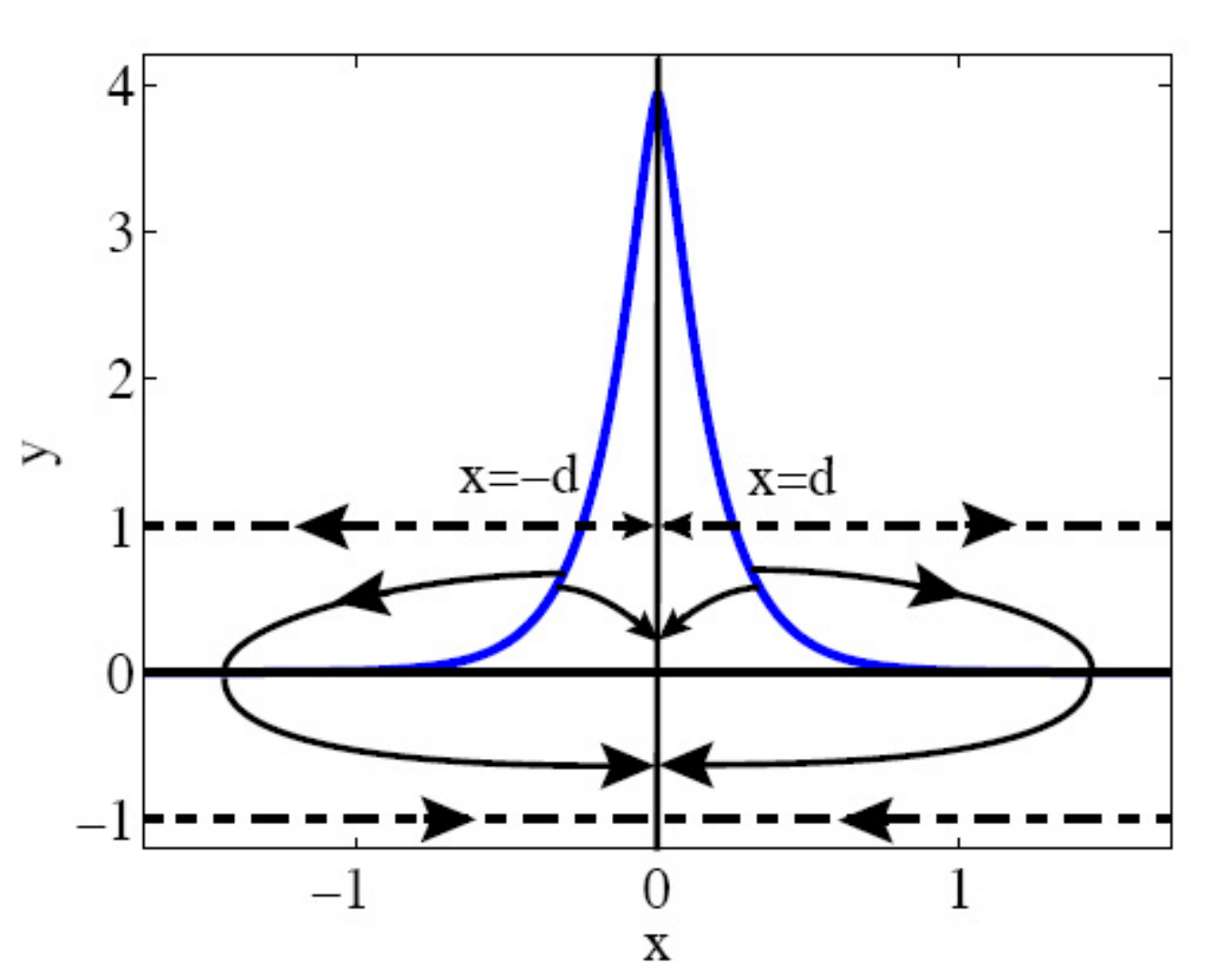}}
}
\caption{(Color online) The nullcline $dx/dt=0$ of the two-clumpon dynamical
system with $\alpha_\rho<\alpha_m$.  The
region contained inside the dotted lines $y=\pm1$ gives the allowed values
of the dynamical variables $\left(x,y\right)$.
 Subfigure~(a)
shows the case when $\beta_m<\alpha_m$.  The lines $x=0$ and $x=\pm\infty$
form the stable equilibria of the system.  All initial conditions flow
 into one of these states; subfigure (b)
 shows the case when $\alpha_m<\beta_m$.  
Initial conditions confined to the  line $y=1$ flow into the fixed points
$\left(0,1\right)$ and $\left(\pm\infty,1\right)$, while all other initial
conditions flow into the line $x=0$. 
}
\label{fig:nullclines2} 
\end{figure}
of all that the $|y|>1$ region of the phase plane is forbidden, since the
$y$-component of the vector field $\left(dx/dt,dy/dt\right)$ vanishes at
$|y|=1$.  The vertical lines $x=0$ and $x=\pm\infty$ are equilibria, although
their stability will depend on the value of the parameters $\left(\alpha_m,\alpha_\rho,\beta_m,B_1,B_2,M\right)$.
The curve across which $dx/dt$ changes sign is called the nullcline.  This
is given by
\[
y=\frac{MH_{\alpha_\rho}'\left(x\right)-B_1H'_{\alpha_m}\left(x\right)H_{\beta_m}\left(x\right)}{B_2H'_{\alpha_m}\left(x\right)},
\]
which on the domain $x\in\left(-\infty,\infty\right)$ takes the form
\[
y=\frac{\alpha_m}{B_2}\left[\frac{M}{\alpha_\rho}e^{-|x|\left(\frac{1}{\alpha_\rho}-\frac{1}{\alpha_m}\right)}-\frac{B_1}{\alpha_m}e^{-\frac{1}{\beta_m}|x|}\right].
\]
Several qualitatively different behaviors are possible, depending on the
magnitude of the values taken by the parameters $\left(\alpha_m,\alpha_\rho,\beta_m,B_1,B_2,M\right)$.
 Here we outline four of these behavior types.
\begin{itemize} 
\item\emph{Case 1:}  The length scales are in the relation $\alpha_m<\alpha_\rho$,
and $\beta_m<\alpha_m$.  The vector field $\left(dx/dt,dy/dt\right)$ and
the nullcline are shown in Fig.~\ref{fig:nullclines}~(a).
There is flow into the fixed points $\left(x,y\right)=\left(\pm{d},1\right)$,
and into the line $x=0$,
while $y$ is a non-decreasing function of time, which follows
from Eq.~\eqref{eq:xtheta_b}.  The ultimate state of the system is thus $x=\pm{d}$,
$\varphi=0$ (alignment), or $x=0$ (merging). In the latter case the final
orientation is given by the integral of Eq.~\eqref{eq:xtheta_b},
\begin{equation}
\tan\left(\frac{\varphi}{2}\right)=\tan\left(\frac{\varphi_0}{2}\right)\exp\left[-B_2\int_0^\infty{dt}\left[H_{\beta_m}\left(x\left(t\right)\right)-H_{\alpha_m}\left(x\left(t\right)\right)\right]\right],\qquad\varphi_0=\varphi\left(t=0\right).
\label{eq:theta_final}
\end{equation}
\item\emph{Case 2:}  The length scales are in the relation $\alpha_m<\alpha_\rho$,
$\alpha_m<\beta_m$.  The vector field and the nullcline are shown in Fig.~\ref{fig:nullclines}~(b).
 All flow not confined to the line $y=1$ is into the line $x=0$, since $y$
 is now a non-increasing function of time.  The ultimate state of the system
 is thus $x=\pm{d}$, $\varphi=0$ (alignment), or $x=0$ (merging). In the
 latter case the final orientation is given by the formula~\eqref{eq:theta_final}.
\item\emph{Case 3:}  The length scales are in the relation $\alpha_\rho<\alpha_m$
and $\beta_m<\alpha_m$.  The vector field and the nullcline are shown in
Fig.~\ref{fig:nullclines2}~(a).
 Inside the region bounded by the line $y=0$ and the nullcline, the flow
 is into the line $x=0$ (merging), and the fixed points
 $\left(\pm{d},1\right)$ are unstable.  The flow below the line $y=0$ is
 towards the line
 $x=0$.  Outside of these regions, however, the flow is into the lines
 $x=\pm\infty$, which shows that for a suitable choice of parameters
 and initial conditions, the clumpons can be made to diverge.
\item\emph{Case 4:}   The length scales are in the relation $\alpha_\rho<\alpha_m$
and $\alpha_m<\beta_m$.  The vector field and the nullcline are shown in
Fig.~\ref{fig:nullclines2}~(b).   The quantity $y$ is a non-increasing function
of time. All flow along the line $y=1$ is directed away from the fixed points
$\left(\pm{d},1\right)$ and is into the fixed points $\left(0,1\right)$,
or $\left(\pm\infty,1\right)$.
 All other initial conditions flow into $x=0$, although initial conditions
 that start above the curve formed by the nullcline flow in an arc and eventually
 reach a fixed point $\left(x=0,y<0\right)$.
\end{itemize}
We summarize the cases we have discussed in Table~\ref{tab:table2_summary}.
Using numerical simulations of Eqs.~\eqref{eq:xtheta}, we have verified that
Cases~(1)--(4) do indeed occur.  The list of cases we have considered is
not exhaustive: depending on the parameters $B_1$, $B_2$, and $M$, other
phase portraits may arise.  Indeed, it is clear from Fig.~\ref{fig:nullclines}
that through saddle-node bifurcations, the fixed points $\left(x,y\right)=\left(\pm{d},1\right)$
may disappear, or additional fixed points $\left(x,y\right)=\left(\pm{d'},-1\right)$
may appear.  Our
\begin{table}[h!b!p!]
\begin{tabular}{|c|c|c|c|c|}
\hline
Case&$\alpha_m$ vs. $\alpha_\rho$&$\alpha_m$ vs. $\beta_m$&Equilibria&Flow\\
\hline
(1)&$\alpha_m<\alpha_\rho$&$\beta_m<\alpha_m$&$\left(x,y\right)=\left(\pm{d},1\right)$;
$x=0$; $x=\pm\infty$&Flow into $x=0$ and $\left(x,y\right)=\left(\pm{d},1\right)$\\
(2)&$\alpha_m<\alpha_\rho$&$\alpha_m<\beta_m$&$\left(x,y\right)=\left(\pm{d},1\right)$;
$x=0$; $x=\pm\infty$&Flow into $x=0$ and $\left(x,y\right)=\left(\pm{d},1\right)$\\
(3)&$\alpha_\rho<\alpha_m$&$\beta_m<\alpha_m$&$\left(x,y\right)=\left(\pm{d},1\right)$;
$x=0$; $x=\pm\infty$&Flow into $x=0$ and $x=\pm\infty$\\
(4)&$\alpha_\rho<\alpha_m$&$\alpha_m<\beta_m$&$\left(x,y\right)=\left(\pm{d},1\right)$;
$x=0$; $x=\pm\infty$&Flow into $x=0$ and $x=\pm\infty$\\
\hline
\end{tabular}
\caption{Summary of the distinct phase portraits of Eq.~\eqref{eq:xtheta}
studied.}
\label{tab:table2_summary}
\end{table}
 analysis shows, however, that it is possible to choose
 a set of parameters $\left(\alpha_\rho,\alpha_m,\beta_m,B_1,B_2,M\right)$
 such that two clumpons either merge, diverge, or are separated by a fixed
 distance.

\subsection*{Numerical Simulations}
To examine the emergence and subsequent interaction of the clumpons, we carry
out numerical simulations of Eq.~\eqref{eq:mag_dens} for a variety
of initial conditions.  We use
an explicit finite-difference algorithm with a small amount of artifical
diffusion.
 We solve the following weak form of Eq.~\eqref{eq:mag_dens}, obtained by
 testing Eq.~\eqref{eq:mag_dens_weak} with $H_{\beta_\Subm}$,
\[
\frac{\partial\overline{\rho}}{\partial{t}}=D_{\mathrm{artif}}\frac{\partial^2\overline{\rho}}{\partial{x^2}}+\int_\Omega{dy}H_{\beta_\Subm}'\left(x-y\right)\rho\left(y,t\right)V\left(y,t\right),
\]
\begin{multline*}
\frac{\partial\mu_i}{\partial{t}}=D_{\mathrm{artif}}\frac{\partial^2\mu_i}{\partial{x^2}}+\int_\Omega{dy}H_{\beta_\Subm}'\left(x-y\right)\Mag_i\left(y,t\right)V\left(y,t\right)\\
+\int_\Omega{dy}H_{\beta_\Subm}\left(x-y\right)\bm{e}_i\cdot\left[\Mag\times\left(\bm{\mu}\times\frac{\delta{E}}{\delta\Mag}\right)\right],
\end{multline*}
where $\overline{\rho}=H_{\beta_m}*\rho$ and $\bm{e}_i$ is the unit vector
in the $i^{\mathrm{th}}$ direction. 
We work on a periodic domain $\Omega=\left[-L/2,L/2\right]$, at a resolution of $250$
gridpoints; going to higher resolution does not noticeably increase
the accuracy of the results.

\begin{figure}[htb]
\subfigure[]{
  \scalebox{0.45}[0.45]{\includegraphics*[viewport=0 0 420 330]{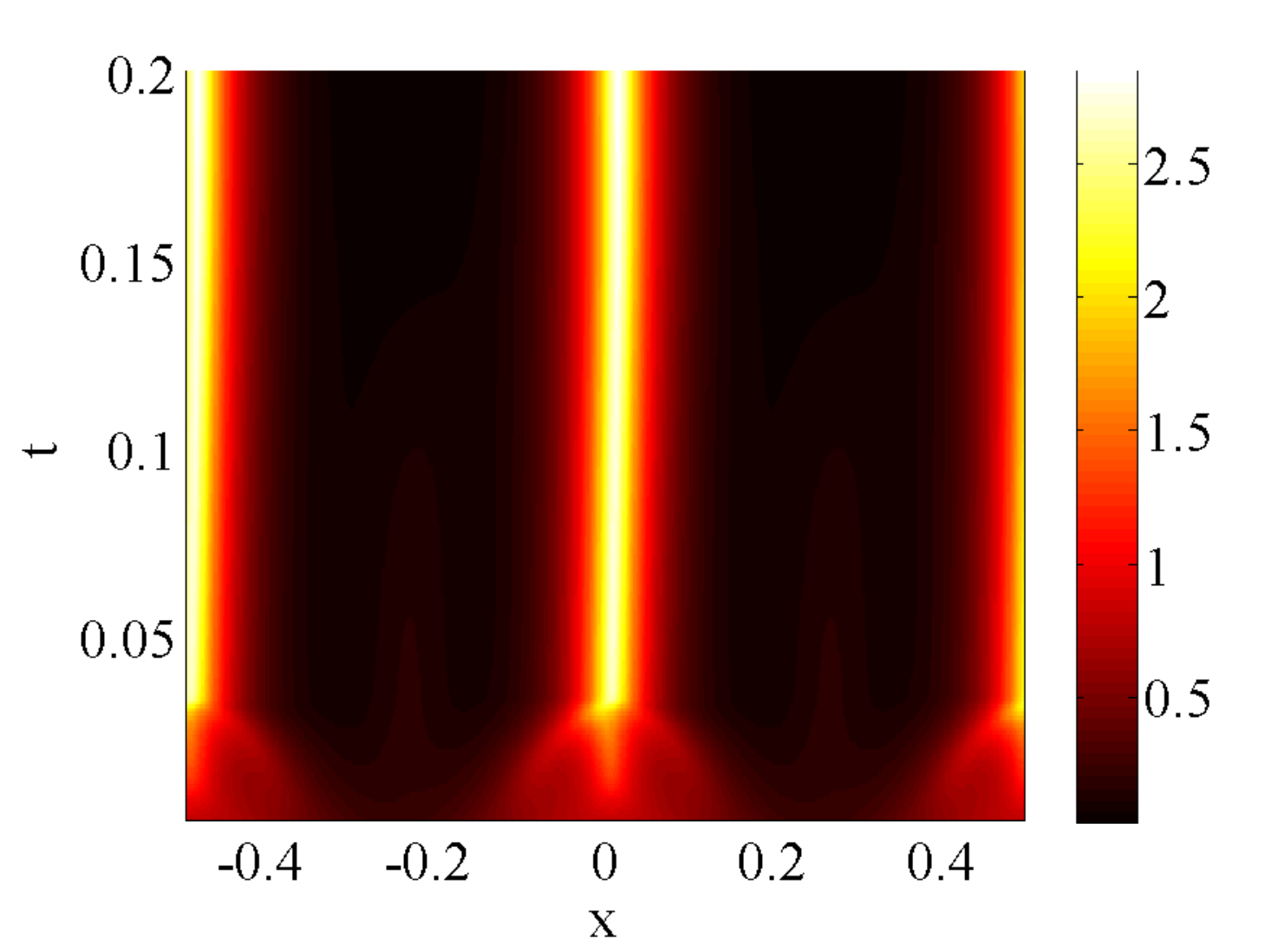}}
}
\subfigure[]{
  \scalebox{0.45}[0.45]{\includegraphics*[viewport=0 0 440 330]{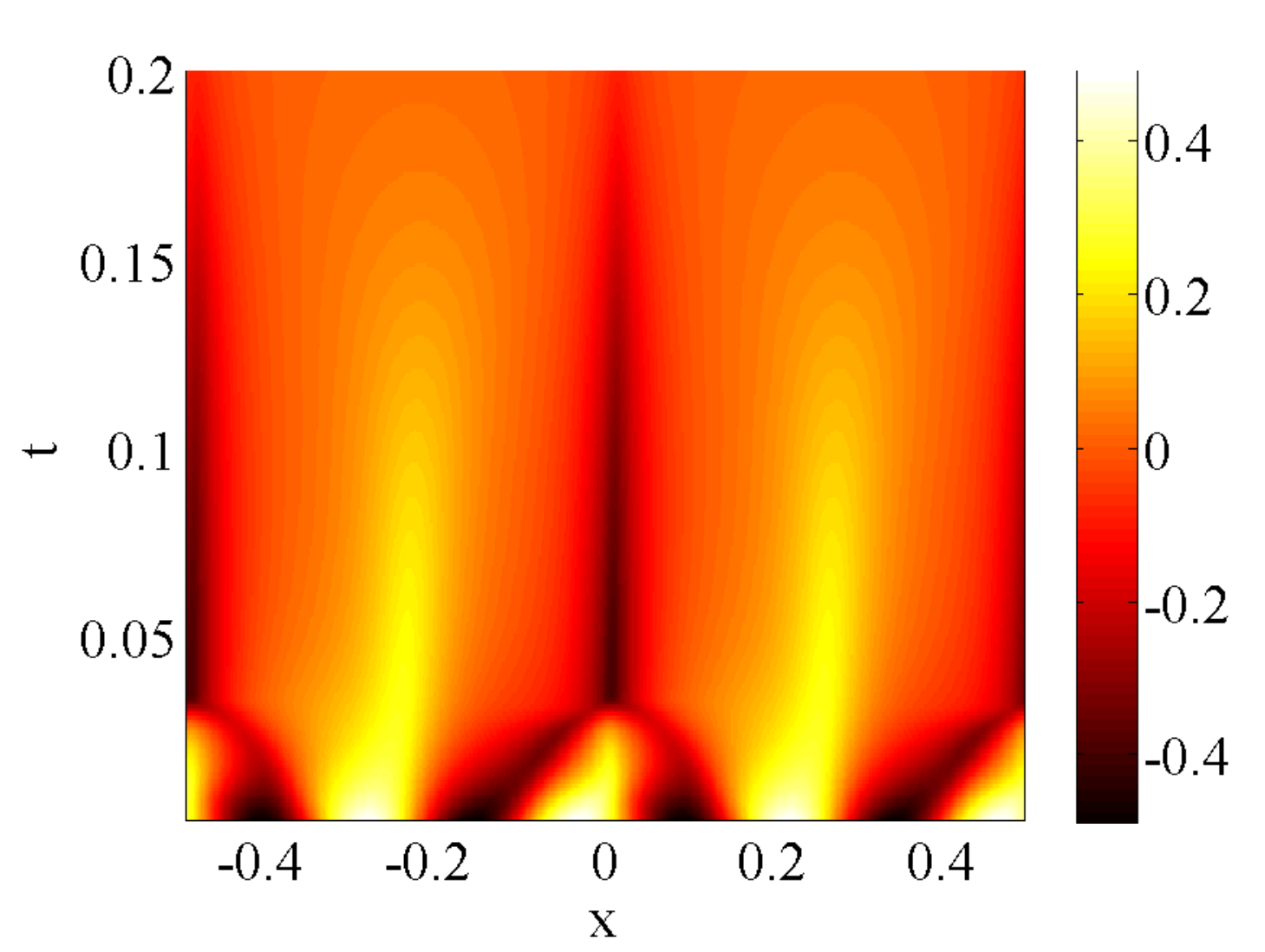}}
}
\caption{(Color online) Evolution of sinusoidally-varying initial conditions
for the density
and magnetization, as in Eq.~\eqref{eq:initial_conditions1}.
 Subfigure~(a) shows the evolution of $H_\rho*\rho$ for $t\in\left[0,0.15\right]$,
 by which time the initial data have formed two clumpons; (b) shows
 the evolution of $\mu_x$.  The profiles of $\mu_y$ and $\mu_z$ are similar.
  Note that
 the peaks in the density profile correspond to the troughs in the magnetization
 profile.  This agrees with the linear stability analysis, wherein disturbances
 in the density give rise to disturbances in the magnetization.
}
\label{fig:evolution_ic2} 
\end{figure}

%
%
The first set of initial conditions we study is the following,
\begin{eqnarray}
\Mag\left(x,0\right)&=&\left(\sin\left(4k_0x+\phi_x\right),\sin\left(4k_0x+\phi_y\right),\sin\left(4k_0x+\phi_z\right)\right),\nonumber\\
\rho\left(x,0\right)&=&0.5+0.35\cos\left(2k_0x\right),
\label{eq:initial_conditions1}
\end{eqnarray}
where $\phi_x$, $\phi_y$, and $\phi_z$ are random phases in the interval
$\left[0,2\pi\right]$, and $k_0=2\pi/L$ is the fundamental wave number.  The
initial conditions for the magnetization vector are chosen to represent the
lack of a preferred direction in the problem.
The time evolution of
equations~\eqref{eq:mag_dens} for this set of initial conditions is shown
in Fig.~\ref{fig:evolution_ic2}.
After a short time, the initial data become singular, and subsequently,
the solution $\left(\rho,\Mag\right)$ can be represented as a sum of clumpons,
\[
\rho\left(x,t\right)=\sum_{i=1}^M a_i\delta\left(x-x_i\left(t\right)\right),\qquad
\Mag\left(x,t\right)=\sum_{i=1}^M \bm{b}_i\left(t\right)\delta\left(x-x_i\left(t\right)\right),\qquad
M=2.
\]
Here $M=2$ is the number of clumpons present at the singularity time.  
This
number corresponds to the number of maxima in the initial density profile.
 The forces exerted by each clumpon on the other balance
 because of the effect of the periodic boundary conditions.  Indeed,
 any number of equally-spaced, identical, interacting particles arranged
 on a ring are in equilibrium, although this equilibrium is unstable for
 an attractive force.  Thus, at late
 times, the clumpons are stationary, while the magnetization vector $\bm{\mu}$
 shows alignment of clumpon magnetizations.
\begin{figure}[htb]
\subfigure[]{
  \scalebox{0.45}[0.45]{\includegraphics*[viewport=0 0 420 330]{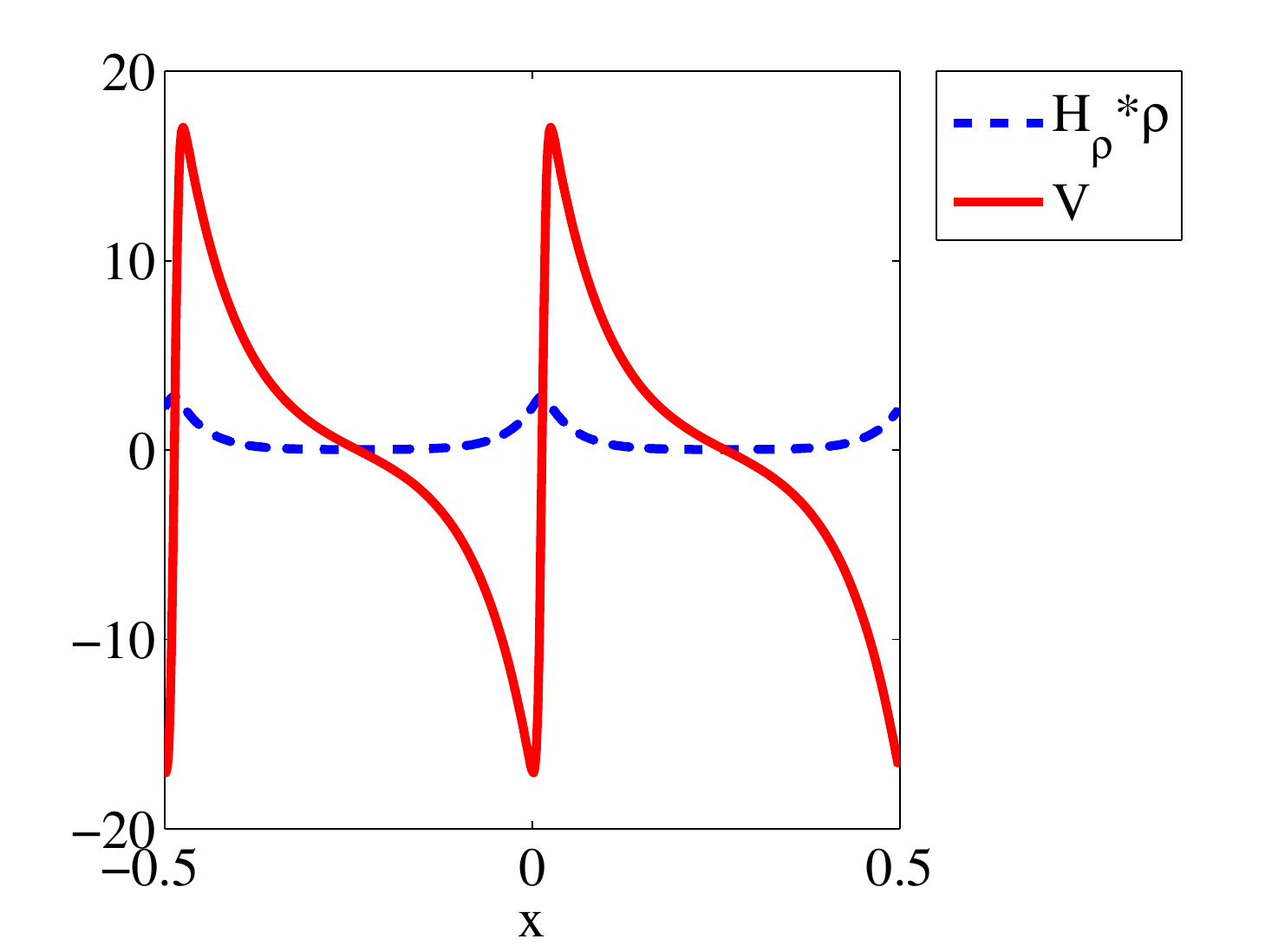}}
}
\subfigure[]{
  \scalebox{0.45}[0.45]{\includegraphics*[viewport=0 0 420 330]{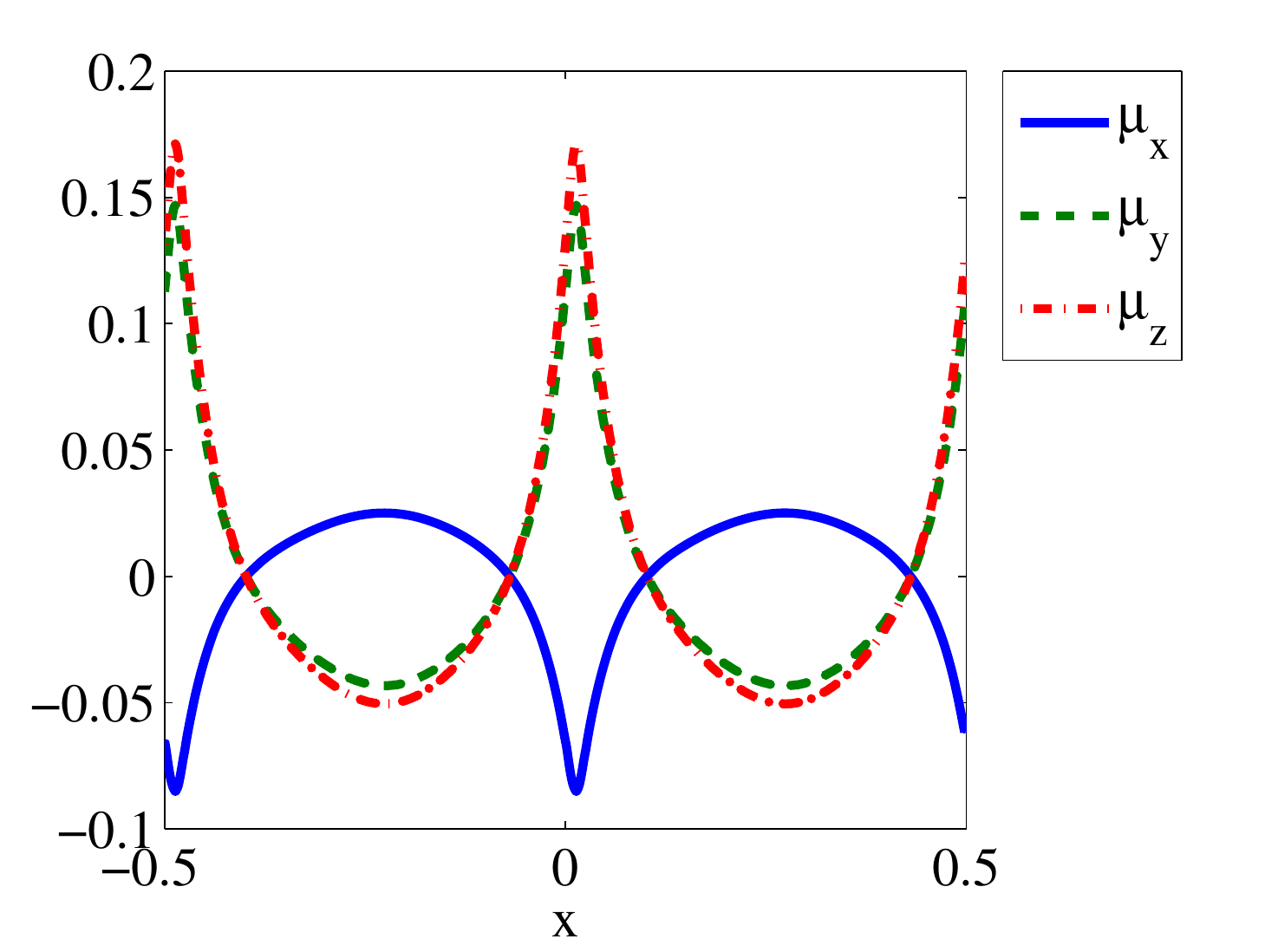}}
}
\caption{(Color online) Evolution of sinusoidally-varying initial conditions
for the density
and magnetization, as in Eq.~\eqref{eq:initial_conditions1}.
 Subfigure~(a) shows the system velocity $V$ given in Eq.~\eqref{eq:velocity},
just before the singularity time; (b) shows the magnetization $\bm{\mu}$
at the same time.  The density maxima emerge at the locations where the convergence
of $-V$ (flow into $x=0$ and $x=\pm L/2$) occurs, and the magnetization develops
extrema there.}
\label{fig:snapshot_ic2} 
\end{figure} 

We gain further understanding of the formation of singular solutions by studying
the system velocity $V$ just before the onset of the singularity.  This is
done in Fig.~\ref{fig:snapshot_ic2}.
Figure~\ref{fig:snapshot_ic2}~(a) shows the development of the two clumpons
from the initial data.  Across each density maximum, the velocity has the
profile $V\approx\lambda\left(t\right)x$, where $\lambda\left(t\right)>0$
is an increasing function of time.  This calls to mind the advection problem
for the scalar $\theta\left(x,t\right)$, studied by Batchelor in the context
of passive-scalar mixing~\cite{Batchelor1959}
\[
\frac{\partial\theta}{\partial t}=\lambda_0x\frac{\partial\theta}{\partial{x}},\qquad
\lambda_0>0.
\]
Given initial data $\theta\left(x,0\right)=\theta_0 e^{-x^2/\ell_0^2}$, the
solution evolves in time as
\[
\theta\left(x,t\right)=\theta_0 e^{-x^2/\left(\ell_0^2e^{-2\lambda_0{t}}\right)},
\]
so that gradients are amplified exponentially in time,
\[
\frac{\partial\theta}{\partial{x}}=-\frac{2\theta_0}{\ell_0^2}xe^{\lambda_0{t}}
e^{-x^2/\left(\ell_0^2e^{-2\lambda_0{t}}\right)},
\]
in a similar manner to the problem studied.  

The evolution of the set of
initial conditions~\eqref{eq:initial_conditions1} has therefore demonstrated
the following:
the local velocity $V$ is such that before the onset of the singularity,
matter
is compressed into regions where $\rho\left(x,0\right)$ is large, to such
an extent that the matter eventually accumulates at isolated points, and
the singular solution emerges.  Moreover, the density maxima, rather than
the magnetization
extrema, drive the formation of singularities.  This is not surprising, given
that the attractive part of the system's energy comes from density variations.
\begin{figure}[htb]
\subfigure[]{
  \scalebox{0.45}[0.45]{\includegraphics*[viewport=0 0 420 330]{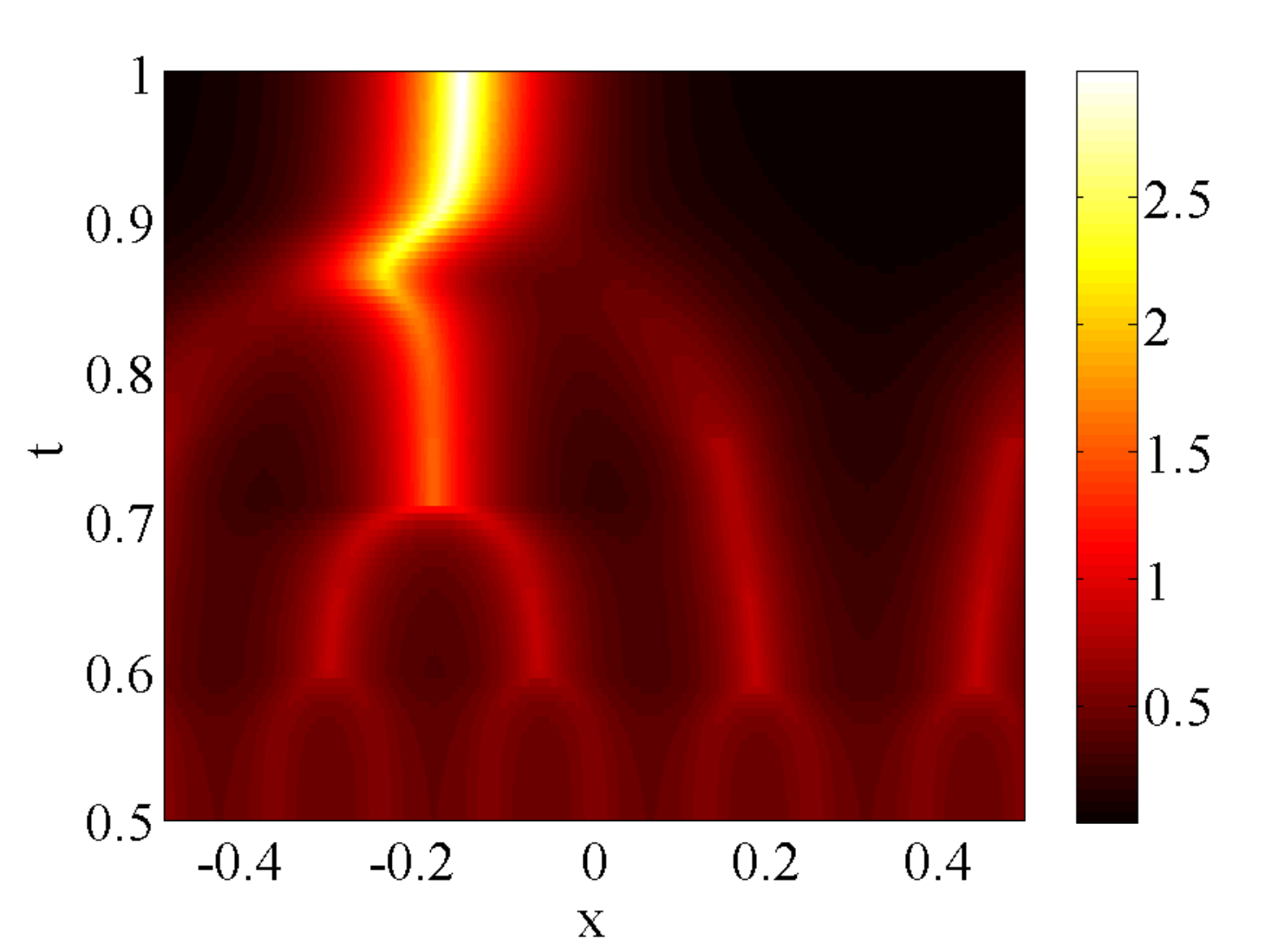}}
}
\subfigure[]{
  \scalebox{0.45}[0.45]{\includegraphics*[viewport=0 0 420 330]{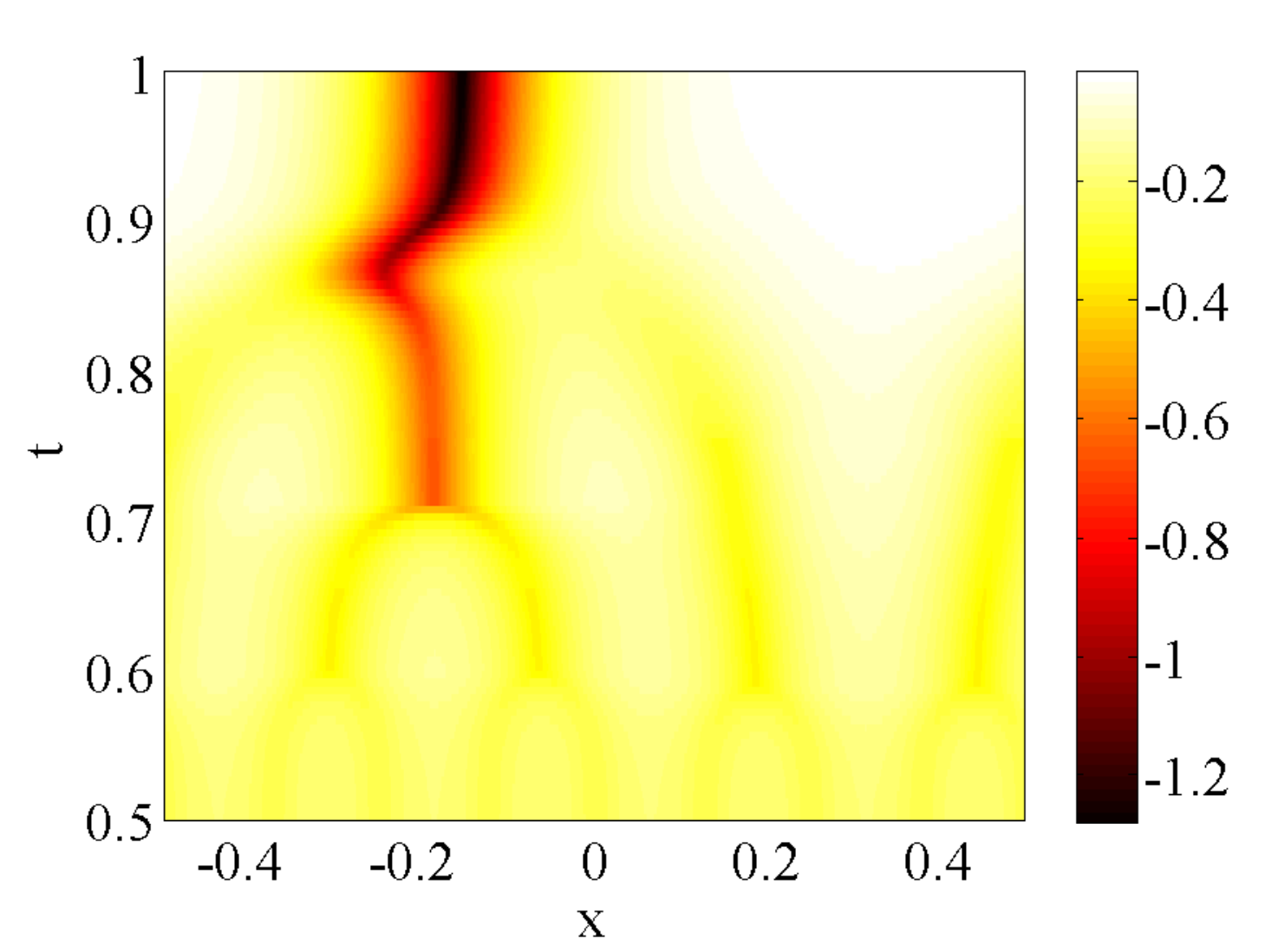}}
}
\caption{(Color online) Evolution of a flat magnetization field and a sinusoidally-varying
density, as in Eq.~\eqref{eq:initial_conditions2}.
 Subfigure~(a) shows the evolution of $H_\rho*\rho$ for $t\in\left[0.5,1\right]$;
 (b) shows the evolution of $\mu_x$.  The profiles of $\mu_y$ and $\mu_z$
 are similar.  At $t=0.5$, the initial data have formed eight equally spaced,
 identical clumpons, corresponding to the eight density maxima in the initial
 configuration.  By impulsively shifting the clumpon
 at $x=0$ by a small amount, the equilibrium is disrupted and the clumpons
 merge repeatedly until only one clumpon remains.
}
\label{fig:evolution_ic1} 
\end{figure}
%
%
%
%
%
%
%
%

\begin{figure}
  \scalebox{0.45}[0.45]{\includegraphics*[viewport=0 0 420 330]{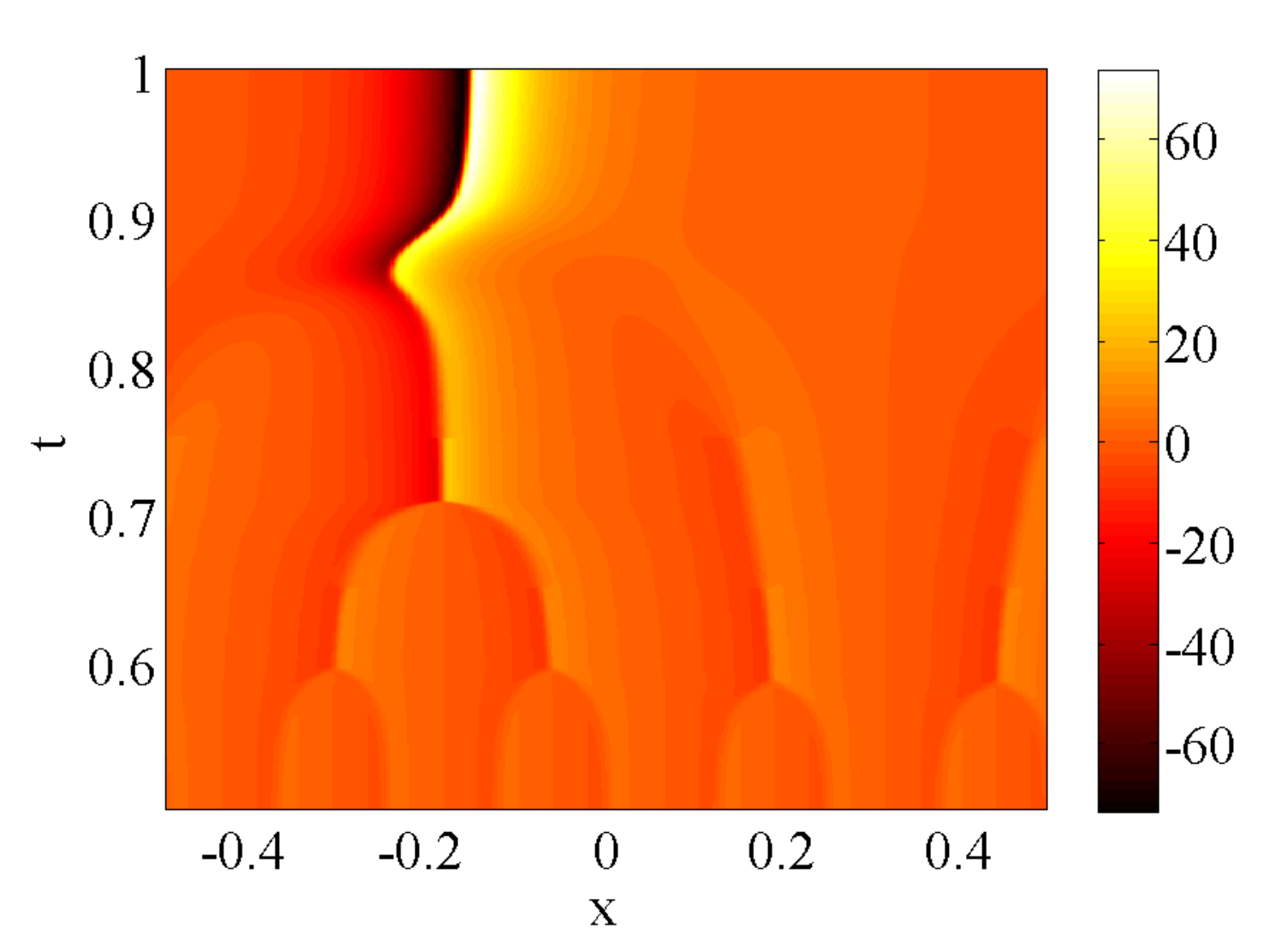}}
\caption{(Color online) Evolution of a flat magnetization field and a sinusoidally-varying
density, as in Eq.~\eqref{eq:initial_conditions2}.
 Shown is the velocity profile for $t\in\left[0.5,1\right]$; the system velocity
 is given by Eq.~\eqref{eq:velocity}.  The velocity $-V$ flows into each
 density
 maximum, concentrating matter at isolated points and precipitating the
 formation of eight equally-spaced identical clumpons.  On a periodic domain,
 such an arrangement is an equilibrium state, although it is unstable.  Thus,
 by impulsively shifting the clumpon at $x=0$ by a small amount, we force
 the clumpons to collapse into larger clumpons, until only a single clumpon
 remains.
 }
\label{fig:V_time_ic1}
\end{figure}

To highlight the interaction between clumpons, we examine the following set
of initial conditions,
\begin{equation}
\Mag\left(x,0\right)=\Mag_0=\text{const.},\qquad
\rho\left(x,0\right)=0.5+0.35\cos\left(8k_0x\right),
\label{eq:initial_conditions2}
\end{equation}
where $k_0=2\pi/L$ is the fundamental wave number.   Since this set of initial
conditions contains a large number of density maxima, we expect a large number
of closely-spaced clumpons to emerge, and this will illuminate the
clumpon interactions.
The time evolution of equations~\eqref{eq:mag_dens} for this set of initial
conditions is shown in Fig.~\ref{fig:evolution_ic1}.
As before, the solution becomes singular after a short time, and is subsequently
represented  by a sum of clumpons,
\[
\rho\left(x,t\right)=\sum_{i=1}^M a_i\delta\left(x-x_i\left(t\right)\right),\qquad
\Mag\left(x,t\right)=\sum_{i=1}^M \bm{b}_i\left(t\right)\delta\left(x-x_i\left(t\right)\right),\qquad
M=8.
\]
Here $M=8$ is the number of clumpons at the singularity time.  This number
corresponds to the number of maxima in the initial density profile.
As before, this configuration of equally spaced, identical clumpons is an
equilibrium state, due to periodic boundary conditions.  Therefore, once
the particle-like state has formed, we impulsively shift the clumpon at $x=0$
by a small amount, and precipitate the merging of clumpons.
The eight clumpons then merge repeatedly until only a single clumpon remains.
%
%
%
%
%
The tendency for the clumpons to merge is explained by the velocity $V$,
which changes sign across a clumpon.  Thus, if a clumpon is within the range
of the force exerted by its neighbours, the local velocity, if unbalanced,
 will advect a given clumpon
\begin{figure}[htb]
\subfigure[]{
  \scalebox{0.45}[0.45]{\includegraphics*[viewport=0 0 420 330]{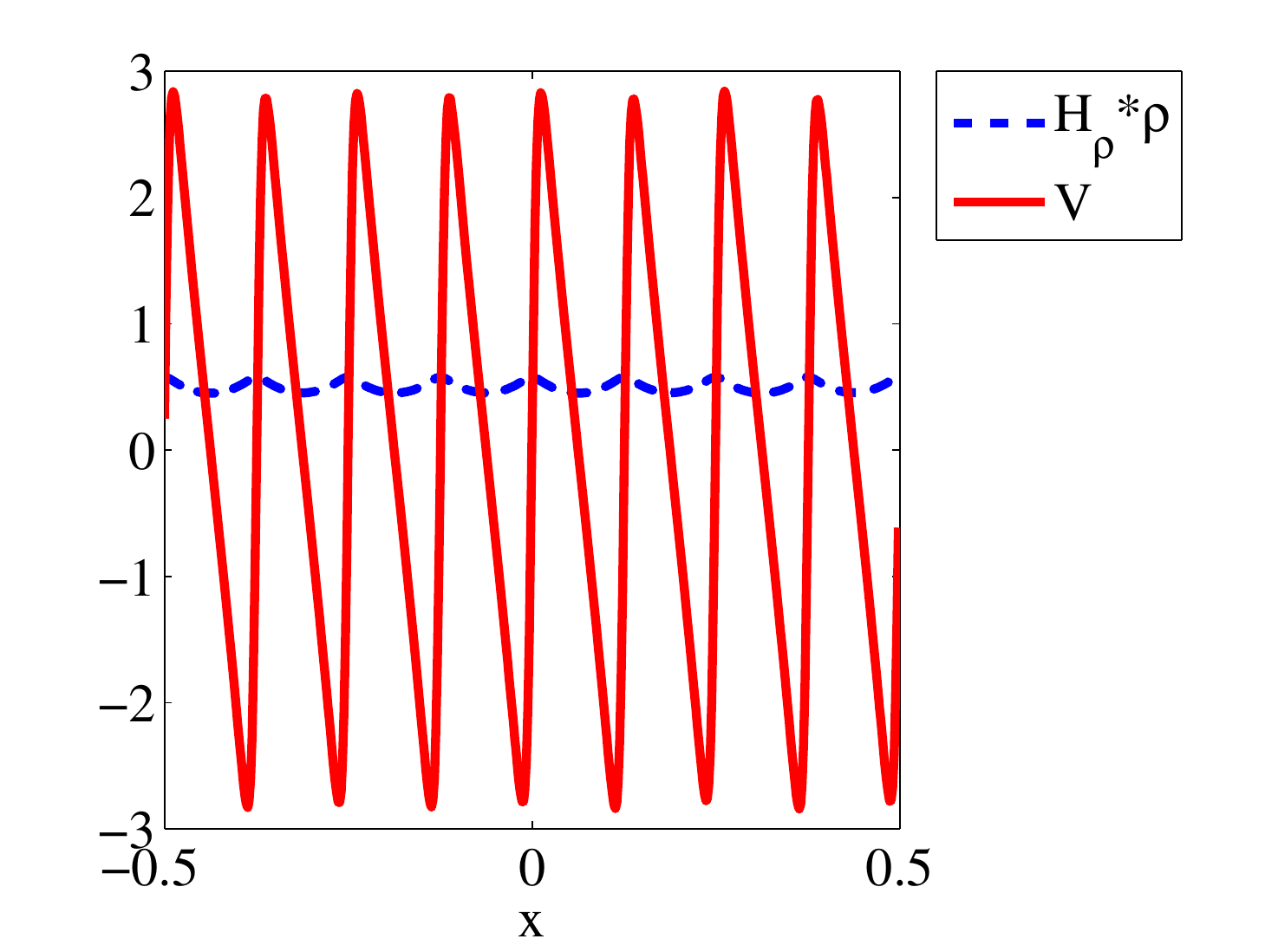}}
}
\subfigure[]{
  \scalebox{0.45}[0.45]{\includegraphics*[viewport=0 0 420 330]{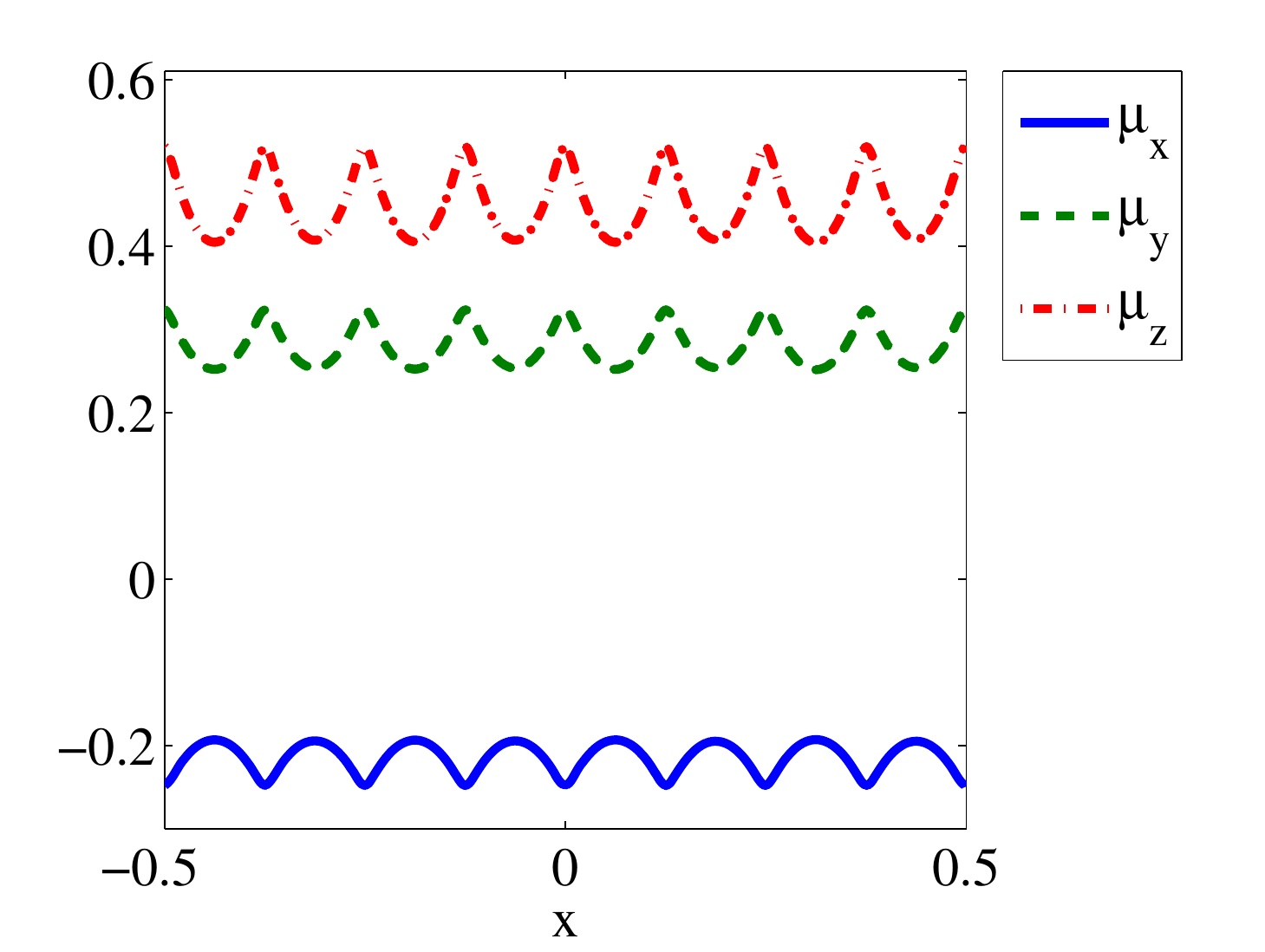}}
}
\caption{(Color online) Evolution of a flat magnetization field and a sinusoidally-varying
density, as in Eq.~\eqref{eq:initial_conditions2}.
 Subfigure~(a) shows the system velocity $V$ given in Eq.~\eqref{eq:velocity}
just before the singularity time; (b) gives the magnetization $\bm{\mu}$
at the same time.  The density maxima emerge at the locations where the convergence
of $-V$ occurs, and the magnetization develops extrema there.}
\label{fig:snapshot_ic1} 
\end{figure} 
in the direction of one of its neighbours, and the clumpons merge.  This
process is shown in Fig.~\ref{fig:V_time_ic1}.  

\section{Conclusions}
\label{sec:conclusions}
We have investigated the non-local Gilbert (NG) equation introduced by Holm,
Putkaradze, and Tronci in~\cite{Darryl_eqn1} using a combination of numerical
simulations and simple analytical arguments.  The NG equation contains two
competing length scales of non-locality: there is a length scale $\alpha$
associated with the range of the interaction potential, and a length scale
$\beta$ that governs the smoothened magnetization vector that appears in
the equation.  When $\alpha<\beta$ all initial configurations of the magnetization
tend to a constant value; while for $\beta<\alpha$ the initial configuration
of the magnetization field develops finer and finer scales.  These two effects
are in balance when $\alpha=\beta$, and the system does not evolve away
from its initial state.  Furthermore, the NG equation conserves the norm
of the magnetization vector $\Mag$, thus providing a pointwise bound on the
solution and preventing the formation of singular solutions.

To study the formation of singular solutions, we couple the NG equation to
a
scalar density equation.  Associated with the scalar density is a negative
energy of attraction that drives the formation of singular solutions and
breaks the pointwise bound on the $\Mag$.  
Three length scales of non-locality
now enter into the problem: the range of the force associated with the scalar
density, the range of the force due to the magnetization,
and the smoothening length.  As before, the competition of length
scales is crucial to the evolution of the system; this is seen in the
linear stability analysis of the coupled equations, in which the relative
magnitude of the length scales determines the stability or otherwise of a
constant state.

Using numerical simulations, we
have demonstrated the emergence of singular solutions from smooth initial
data, and have explained this behavior by the negative energy of attraction
produced by the scalar density.  The singular solution consists of a weighted
sum of delta functions, given in Eq.~\eqref{eq:clumpon_sln},
which we interpret 
as interacting particles or clumpons.  The clumpons evolve under simple finite-dimensional
dynamics.  We have shown that a system of two clumpons is governed by a two-dimensional
dynamical system that has a multiplicity of steady states.  Depending on
the length scales of non-locality and the clumpon weights, the two clumpons
can merge, diverge, or align and remain separated by a fixed distance.
  
Our paper thus gives a qualitative description of the dynamics.  Future work
will focus on the regularity of solutions
of the NG equation, and the existence and regularity of solutions for the
coupled density-magnetization equations.  Bertozzi and Laurent~\cite{Bertozzi2007}
have studied the simpler (uncoupled) non-local scalar density equation, proving
existence, uniqueness, and blowup results using techniques from functional
analysis, and a similar analysis will illuminate the equations we have studied.
The behavior of singular solutions in higher dimensions is another topic
that deserves further study.

DDH was partially supported by the US Department of Energy, Office of Science,
Applied Mathematical Research and the Royal Society Wolfson Research Merit
Award.  CT was also partially supported by the Royal Society Wolfson Research
Merit Award.  L.\'O.N. was supported by the Irish government and the UK Engineering
and Physical Sciences Research Council.

%

%
%
%
\end{document}